\newcommand {\reff}	{r_{\rm eff}}	
\newcommand {\FeH} 	{{\rm [Fe/H]}}	
\newcommand {\OFe} 	{{\rm [O/Fe]}}	
\newcommand {\V}	{{\sl V\/}}	
\newcommand {\I}	{{\sl I\/}}	
\newcommand {\eg} 	{e.g.\ }	
\newcommand {\etal}	{et al.\ }	
\newcommand {\sect}[1]	{Sec.\ #1}	

\documentstyle[10pt,aaspp4]{article}
\tighten
  
\received{}
\revised{}
\accepted{}
\journalid{}{}
\articleid{}{}
\paperid{}

\lefthead{Holland \etal}
\righthead{G302 \& G312}

\slugcomment{to appear in {\sl AJ}, October 1997}

\begin{document}

\title{Deep {\sl HST\/} \V- and \I-Band Observations of Two Globular
Clusters  in the Halo of  M31\footnote{Based  in observations with the
NASA/ESA {\sl Hubble Space Telescope}, obtained at the Space Telescope
Science  Institute,   which  is   operated   by  the   Association  of
Universities  for Research in  Astronomy,  Inc.,  under NASA  contract
NAS5-26555.}}

\author{Stephen Holland, Gregory G. Fahlman, \& Harvey B. Richer}

\affil{Department of Physics \& Astronomy \\
\#129--2219 Main Mall \\
University of British Columbia \\
Vancouver, B.C., Canada \\
V6T 1Z4}

\begin{abstract}
	We present deep ($V \simeq 27$) \V- and \I-band stellar
photometry of G302 and G312, two globular star clusters in the halo of
M31.  These data were obtained using the {\sl Hubble Space
Telescope\/}'s Wide Field/Planetary Camera 2.  We find iron abundances
of $\FeH = -1.85 \pm 0.12$ for G302 and $\FeH = -0.56 \pm 0.03$ for
G312, consistent with spectroscopic measurements.  The
color--magnitude diagrams for each cluster show no evidence for an
intermediate-aged population of stars, or a second parameter effect in
the morphology of the horizontal branch.  G302 shows no evidence for a
color gradient but the inner regions of G312 are bluer than the outer
regions.  G312 shows no evidence of ellipticity or an extended halo of
unbound stars.  G302 has a projected ellipticity of $\epsilon = 0.195
\pm 0.012$ with the projected major axis oriented towards the center
of M31.  G302 also shows evidence of an extended asymmetric stellar
halo extending to at least twice the fitted Michie--King tidal radius.
The amount of mass beyond the tidal radius of G302 is consistent with
the stellar escape rates which have been predicted by $N$-body
simulations of globular clusters in the Galactic tidal field.
\end{abstract}

\keywords{galaxies: individual (M31) ---
galaxies: star clusters ---
stars: luminosity function
}


\section{Introduction\label{SECTION:intro}}

	Globular star clusters (GCs) are pressure-supported
collections of between $\sim 10^4$ and $\sim 10^6$ stars which are
usually associated with galaxies, although there is evidence that some
clusters of galaxies contain a population of ``free'' GCs which are
associated with the cluster potential as a whole and not any
individual galaxy (West \etal \markcite{WC95}1995).  Individual
Galactic GCs are made up of stars with a single overall chemical
abundance suggesting that they formed in a single star formation
event.  GCs typically have integrated magnitudes of $-10 \le V \le -4$
making GC systems visible out to redshifts of $z \sim 0.04$, the
approximate distance to the Great Wall galaxies.  The shape of the GC
luminosity function (LF) has been assumed to be universal, so the GC
LF has been used as a distance indicator (see Harris
\markcite{H91}1991 for a review).  However, recent work has suggested
that the shape of the GC LF may depend on the metallicity of the GC
system (Ashman \etal \markcite{AC95}1995) and the details of stellar
evaporation from individual GCs (Okazaki \& Tosa \etal
\markcite{OT95}1995).  Because of the wide-spread use of GCs to
determine distances to external galaxies, it is important to determine
if GCs truly are the same from one galaxy to the next.  This is best
determined by studying the physical structures and stellar populations
of GCs in nearby galaxies.

	The nearest large GC system outside the Milky Way Galaxy is
that of the Andromeda Galaxy ($=$ M31 $=$ NGC 224).  M31 is located at
a distance of 725 kpc ($\mu_0 = 24.3$, van den Bergh
\markcite{V91}1991) so individual stars in M31's GCs can be easily
resolved with the {\sl Hubble Space Telescope\/} ({\sl HST\/}) and
large ground-based telescopes at sites with sub-arcsecond seeing.

	M31's low inclination ($i = 12\fdg5$, Hodge
\markcite{H92}1992) means that many of its GCs are not superimposed
against the disk of M31, making identification of GCs, and photometry
of their stars, relatively straightforward.  M31 offers a unique
laboratory to study the outer regions of GCs.  Because of their small
fields-of-view it is not practical to used CCDs to obtain star counts
in the outer regions of Galactic GCs.  Beyond distances of
approximately half the tidal radius, $r_t$, the projected stellar
densities of the clusters are overwhelmed by random fluctuations in
the background stellar number density (Innanen \etal
\markcite{IH83}1983).  However, M31's GCs have sufficiently small
angular sizes ($\theta \sim 5\arcsec$ to $15\arcsec$) that both the
cluster and the background can be imaged on a single large format CCD
image.  This eliminates the need to match photometric zero-points
between the cluster and the background, which makes possible a more
precise subtraction of the background light from the cluster light.

        The first study of the internal structures of M31 GCs was
undertaken by Battistini \etal \markcite{BB82}(1982), who estimated
core radii for several clusters.  Pritchet \& van den Bergh
\markcite{PV84}(1984) found that the surface brightness profile 
for G1 ($=$ Mayall II; the G-numbers used in this paper are from
Sargent \etal \markcite{SK77}[1977]) had an excess of light at large
radii compared to the best-fitting seeing-convolved analytical King
\markcite{K66}(1966) model.  They did find that G1 was well fit by
empirical King \markcite{K62}(1962) models with core radii of $r_c
\lesssim 0.5$ pc.  Crampton \etal \markcite{CC85}(1985) used
seeing-convolved King \markcite{K62}(1962) models to derive core radii
for nearly 500 M31 GCs.  Their values, however, are systematically
$\sim$50\% larger than those of Battistini \etal
\markcite{BB82}(1982), despite the better seeing conditions of the
Crampton \etal \markcite{CC85}(1985) data set.  Bendinelli
\etal \markcite{BP90}(1990) used ground-based data to produce
seeing-deconvolved radial profiles for six bright GCs in M31, but
seeing and pixel scale limitations restricted them to resolutions of
$\sim$0\farcs3, insufficient to resolve the cores of the clusters.
Still, their data suggest that M31 GCs have King-like profiles similar
to those of Galactic GCs.

        The core structures of some of M31's GCs have been studied
using the pre-refurbished {\sl HST}.  Bendinelli \etal
\markcite{BC93}(1993) detected a power-law density cusp in G105 using
{\sl HST\/}'s Faint Object Camera (FOC) images and a variety of image
restoration and seeing deconvolution techniques.  In addition, Fusi
Pecci \etal \markcite{FB94}(1994) used similar methods to obtain
half-width at half-maxima (HWHM) and half-light radii ($r_h$) for
thirteen M31 GCs from FOC images.  They found HWHMs similar to the
core radii of Galactic GCs.  Their data, however, could not be used to
find the tidal radii of these clusters because the point-spread
function (PSF) before the refurbishment of the {\sl HST\/} overfilled
the FOC's field of view.

	Cohen \& Freeman \markcite{CF91}(1991) derived tidal radii for
thirty M31 GCs by fitting seeing-convolved King models.  Although
their fits to individual clusters were quite uncertain they did find a
mean tidal radius for the M31 clusters---after adjustment for
differences in galactic masses and rotation velocities---which was
very similar to that of the Milky Way clusters.  GCs do not exist in
isolation but sit in the tidal field of a galaxy.  Any stars that
evaporate from a GC by the galaxy will have velocity vectors similar
to the velocity vector of the GC\@.  This can result in the GC being
surrounded by an extended halo of unbounded stars which move in
approximately the same direction as the GC and have approximately the
same velocity.  This idea has been explored numerically by Oh \& Lin
\markcite{OL92}(1992) who predicted that GCs could be surrounded by
extended halos of escaped stars which can persist for up to a Hubble
time.  The size and shape of such a halo can be influenced by the
tidal field of the parent galaxy.  Evidence for extended halos has
been observed in some Galactic GCs by Grillmair
\etal \markcite{GF95}(1995).  In addition, Grillmair \etal
\markcite{GA96}(1996, hereafter referred to as GAF) have observed an
excess of resolved and unresolved stars beyond the formal Michie--King
(MK) tidal radii of several GCs in M31, as would be expected if
extended halos were present.

        There has been some interest in determining the ellipticities
of M31 GCs.  Pritchet \& van den Bergh \markcite{PV84}(1984) measured
an ellipticity of $\epsilon = 0.22$ for the region of G1 with
$12\arcsec \lesssim r \lesssim 35\arcsec$.  Spassova \etal
\markcite{S88}(1988) measured ellipticities for approximately two dozen
GCs while a study by Lupton \markcite{L89}(1989) suggests that the
mean ellipticity measured in the inner 7 to 14 pc of a M31 GC
($\overline{\epsilon} = 0.08$) is indistinguishable from the mean
ellipticity of Galactic GCs.  In the outer 14 to 21 pc, however, the
mean ellipticity of an M31 cluster is $0.11 \pm 0.08$, greater than
that of the Galactic GCs ($\overline{\epsilon} = 0.08 \pm 0.07$) and
similar to GCs in the Large Magellanic Cloud ($\overline{\epsilon} =
0.11 \pm 0.07$).  Baev \etal \markcite{BS97}(1997) found systematic
differences between the shapes of M31's disk and halo GCs.  They found
that the disk GCs are triaxial ellipsoids while the halo GCs are
oblate or prolate spheroids, but cautioned that this is a preliminary
result since the sample of halo GCs in their study is small compared
to the sample of disk GCs.
 
	The first color--magnitude diagrams (CMDs) for GCs in M31 were
for G1 by Heasley \etal \markcite{HC88}(1988) and G219 ($=$ Mayall IV)
by Christian \& Heasley \markcite{CH91}(1991).  Couture \etal
\markcite{CR95}(1995) undertook a systematic study of five M31 GCs
(G11, G319, G323, G327 $=$ Mayall VI, and G352 $=$ Mayall V) with a
range of iron abundances.  Unfortunately none of these ground-based
studies were able to reach the level of the horizontal branch (HB) at
$V \simeq 25$.  The first CMDs constructed from {\sl HST\/} data were
for G1 (Rich \etal \markcite{RM96}1996); G58, G105, G108, and G219
(Ajhar \etal \markcite{AG96}1996); and G280, G351, and Bo468 (Fusi
Pecci \etal \markcite{FP96}1996).  These CMDs were able to resolve
stars one to two magnitudes below the red portion of the HB\@.

	In order to study the stellar populations and structures of
GCs in M31 we obtained deep {\sl HST\/} Wide-Field/Planetary Camera-2
(WFPC2) images of two halo GCs (G302 and G312) located $32\farcs1$ and
$49\farcs8$ respectively from the center of M31 along the southeast
minor axis.  Our data is unique in that we have centered the GCs on
the WF3 CCD in order to take advantage of the WFC's field of view to
search for extended halos around each GC\@.


\section{Observations and Data Reductions\label{SECTION:obs_data}}

\subsection{Observations\label{SECTION:obs}}

	We obtained deep \V- and \I-band images of two bright GCs,
G302 ($\alpha_{2000.0} = 00^{\rm h} 45^{\rm m} 25\fs2,\,\,
\delta_{2000.0} = +41\arcdeg 05\arcmin 30\arcsec$) and G312
($\alpha_{2000.0} = 00^{\rm h} 45^{\rm m} 58\fs8,\,\,\delta_{2000.0} =
+40\arcdeg 42\arcmin 32\arcsec$) in the halo of M31.  These data were
obtained using the {\sl HST\/} in cycle 5 program \#5609.  We used the
WFPC2 operating at $-88\arcdeg$C with a gain setting of 7 ${\rm
e}^-$/ADU.

	Table~\ref{TABLE:obs_log} lists the exposures obtained in each
filter.  Exposure times were based on the central surface brightnesses
of the two GCs as determined from ground-based observations (Holland
\etal \markcite{HF93}1993, Holland \markcite{H97}1997) and chosen
to avoid saturating the cores of the GCs.  The total exposure times
were 4320 seconds in the F555W (WFPC2 broadband \V) filter and 4060
seconds in the F814W (WFPC2 broadband \I) filter for each field.  The
data was pre-processed through the standard STScI pipeline for WFPC2
data.  Known bad pixels were masked, and geometric corrections were
applied using standard techniques.

	G302 is located $32\farcm1$ from the center of M31 (a
projected distance of $R_{\rm M31} = 6.8$ kpc) approximately along the
southeast minor axis while G312 is located $49\farcm8$ from the center
of M31 ($R_{\rm M31} = 10.5$ kpc) approximately along the southeast
minor axis.  Each GC was centered on the WF3 CCD\@.  We chose to
locate the GCs on the WF3 CCD, instead of taking advantage of the
higher spatial resolution offered by the smaller pixels of the PC CCD,
for three reasons:

	First, the tidal radii of G302 and G312, as determined by
fitting MK models (Michie \markcite{M63}1963, King \markcite{K66}1966)
to the surface brightness profiles of G302 and G312 derived from HRCam
observations taken at the CFHT, are $r_t \simeq10\arcsec$ (Holland
\etal \markcite{HF93}1993, Holland\markcite{H97} 1997).  Therefore,
each GC would occupy a significant fraction of the total field of view
of the PC, leaving insufficient area to make reliable determinations
of the spatial variations in the background stellar density.

	Second, Grillmair \etal \markcite{GF95}(1995) found evidence
for extended stellar halos, extending to $\sim$ 2 to 3 times the tidal
radius, in several Galactic GCs.  Recently, GAF and Holland
\markcite{H97}(1997) found weak evidence for stellar density
enhancements out to $\sim 2 r_t$ in some M31 GCs including G302 and
G312.  By centering the GCs on the WF3 CCD we ensured that as much as
possible of any extended halos would be located on the same CCD as the
GC itself.  This enabled us to avoid systematic effects that may occur
if surface brightness and stellar number density data are transferred
between the PC and WFC CCDs, with their different pixel scales and
sensitivities.

	Third, by locating the GCs at the center of the WF3 CCD we
were able to use the WF2 and WF4 CCDs to probe the stellar populations
in the halo of M31.  These results are published in Holland \etal
\markcite{HF96}(1996, hereafter referred to as HFR).  Having two
background fields for each GC field enables us to differentiate
between stellar number-density enhancements that are due to
statistical fluctuations in the star counts from M31 halo stars, and
those that are due to physical structures, such as extended halos of
unbound stars, that are related to the GCs.

\subsection{Data Reductions\label{SECTION:data}}

	Photometry was performed on stars imaged by the WFPC2's Wide
Field CCDs using the {\sc daophot ii}/{\sc allframe} software (Stetson
\markcite{S87}1987, Stetson \markcite{S94}1994).  All nineteen images
for each field were re-registered to a common coordinate system and
median-combined.  This eliminated cosmic ray hits and increased the
signal-to-noise ratio (S/N) in our data allowing fainter stars to be
detected.  In order to eliminate large-scale gradients in the images
due to unresolved light from the GCs (the gradient in the light from
the halo of M31 was $\lesssim$ 1\% in both the G302 and G312 fields) a
square median-filter was run over the combined image to produce a map
of the large-scale gradients.  This smoothed image was then subtracted
from the combined image and a constant sky added back to produce a
median-subtracted image.  The size of the square median-filter was
$\sim$5 times the stellar full-width at half-maximum (FWHM) to avoid
smoothing out structure in the stellar images (e.g. Stetson \& Harris
\markcite{SH88}1988).  Subtracting the median-filtered image resulted
in a reasonably clean subtraction of the integrated light from each GC
to within $\sim$ 25 pixels ($\simeq$ 2\farcs5 $\simeq$ 9 pc) of the
center of each GC\@.  This process is the digital equivalent of
unsharp masking.

	The {\sc daophot find} routine, with a {\sc find} threshold of
$7.5\sigma_{\rm sky}$, was used to detect peaks on the
median-subtracted image.  We experimented with different {\sc find}
thresholds and found that $7.5\sigma_{\rm sky}$ excluded most of the
mis-identifications of noise spikes and cosmic-ray events as stars at
faint magnitudes while including almost all of the stars to the
photometric limit of the data.  {\sc Find} was able to reliably detect
stars to within $\sim$ 25 pixels ($\sim$ 2\farcs5) of the center of
each GC\@.  Artificial star tests showed that detections closer to the
centers of the GCs than this were usually spurious.

	The resulting list of stellar candidates was used as input to
{\sc allframe}.  {\sc Allframe} does simultaneous PSF fitting on all
the {\sl original\/} images ({\sl not\/} the combined
median-subtracted image) to preserve the photometric properties of the
data.  The PSFs used (Stetson \markcite{S96}1996) were Moffatians
(Moffat \markcite{M69}1969) with $\beta = 1.5$ and a look-up table of
residuals.  The PSFs varied quadratically over each CCD to account for
variations in the form of the PSF from one part of the CCD to another.
Only stars which appeared in at least seven frames for each filter
were considered to be real stars.  Tests with different minimum
numbers of frames showed that this requirement provides a clean CMD
without significantly affecting the limiting magnitude.

	Aperture corrections and photometric calibrations were
performed in exactly the same manner as described in HFR\@.  We
adopted a distance modulus of $\mu_0 = (m-M)_0 = 24.3 \pm 0.1$, a
reddening of $E_{V-I} = 0.10 \pm 0.03$, and interstellar extinctions
from Tables 12A and 12B of Holtzman \etal \markcite{HB95}(1995).
These extinction values are dependent on the stellar spectrum.  For a
K5 giant the extinctions were $A_V = 0.320$ and $A_I = 0.190$.  See
HFR for a discussion of why these values were adopted.
Table~\ref{TABLE:apcor} lists the adopted aperture corrections for the
WFPC2 CCDs.  We were unable to obtain reliable aperture corrections
for the PC images of the G312 field due to a lack of bright stars in
this field.  Photometry in this field exhibited an unusually high
degree of scatter, although the stellar images appeared to be normal
to the eye.  As a result, we have not included this data in any of the
analysis presented here.

	{\sc Allframe}'s $\chi$ statistic represents the ratio of the
observed pixel-to-pixel scatter in the PSF fitting residuals to the
expected amount of scatter given the noise properties of the data.
Values of $\chi$ which are significantly greater than unity indicate
that the image was not well fit by the stellar PSF and thus may not be
a star.  In light of this we discarded any stars with a final $\chi$
value of greater than 2.  In order to eliminate stars with poorly
determined magnitudes we used an interactive approach based on the
uncertainty in the calibrated magnitude.  A plot of uncertainty vs.\
magnitude was made and a locus defined which corresponded to the
expected photometric scatter at a given magnitude.  Any stars which
were judged to have a significantly greater uncertainty than the
typical value for that magnitude were discarded.  Visual examinations
of the CMDs and the WF images suggested that we did not remove any
legitimate stars from our sample.  We wish to stress that this culling
was done {\sl solely\/} on the basis of the photometric uncertainty of
each star, not on the star's location on the CMD or the CCD\@.  Mean
photometric uncertainties for our entire culled data set are listed in
Table 3 of HFR\@.  The mean photometric uncertainties, based on the
formal photometric uncertainties returned by {\sc daophot
ii/allframe}, for the GC stars are listed in
Table~\ref{TABLE:mag_errs} of this paper.

	The complete set of calibrated stellar photometry for G302 and
the surrounding fields is listed in
Table~\ref{TABLE:G302_sample_phot}.  The photometry for G312 and its
surrounding fields (with the exception of the PC field) is listed in
Table~\ref{TABLE:G312_sample_phot}.  Column (1) gives the CCD the star
was observed on.  Column (2) gives an identification number for each
star.  The identification numbers are only unique for stars on a given
CCD\@.  Columns (3) and (4) give the location of the star on the CCD
in pixel coordinates.  Column (5) is the \V-band magnitude, column (6)
the 1-$\sigma$ uncertainty in the \V-band magnitude, and column (7) is
the value of the {\sc daophot} $\chi$ statistics.  Columns (8) to (10)
are the same as columns (5) to (7) except they give values for the
\I-band.  Tables~\ref{TABLE:G302_GC_sample_phot}~and~\ref{TABLE:G312_GC_sample_phot} list the photometry for stars with $r
\le 10\arcsec$ of the center of G302 and G312. The data is the same as
for
Tables~\ref{TABLE:G302_sample_phot}~and~\ref{TABLE:G312_sample_phot}
except an extra column, giving the distance of the star from the
center of the GC, has been added.  Only the first ten lines of
Tables~\ref{TABLE:G302_sample_phot},~\ref{TABLE:G312_sample_phot},~\ref{TABLE:G302_GC_sample_phot},~and~\ref{TABLE:G312_GC_sample_phot}
are published here.  The full tables can be found on the AAS CD-ROM
Volume 9.  A total of 12,289 stars were found in the three G302 WFC
fields while 3,355 stars were found in the three G312 WFC fields.
There are 626 stars within $10\arcsec$ of the center of G302 and 325
stars within $10\arcsec$ of the center of G312.

\subsection{Artificial Star Tests\label{SECTION:art_stars}}

	In order to determine the effects of crowding, as well as the
efficiencies of our star-finding and photometry techniques, on our
photometry, we undertook a series of artificial star tests.  We used
the following procedure to generate and recover artificial stars:

\begin{enumerate}

\item
For each field we generated a series of artificial stars with
$(V,(V-I))$ colors based on the observed CMD and LF of the M31 halo
stars.

\item
The artificial stars were added to each of the original images for a
field.  The total number of artificial stars added in each field was
$\sim 10$\% of the total number of stars in that field.

\item
The entire photometry process described in \sect~\ref{SECTION:data}
({\sc find}/{\sc allframe}/culling/matching) was performed on this new
set of images and the resulting photometry was matched against the
original list of artificial stars for that field.

\item
This process was repeated until we had recovered approximately the
same number of artificial stars as there were real stars found on that
CCD\@.

\end{enumerate}

\noindent
This process was performed for each CCD field.

\begin{figure}
\plotone{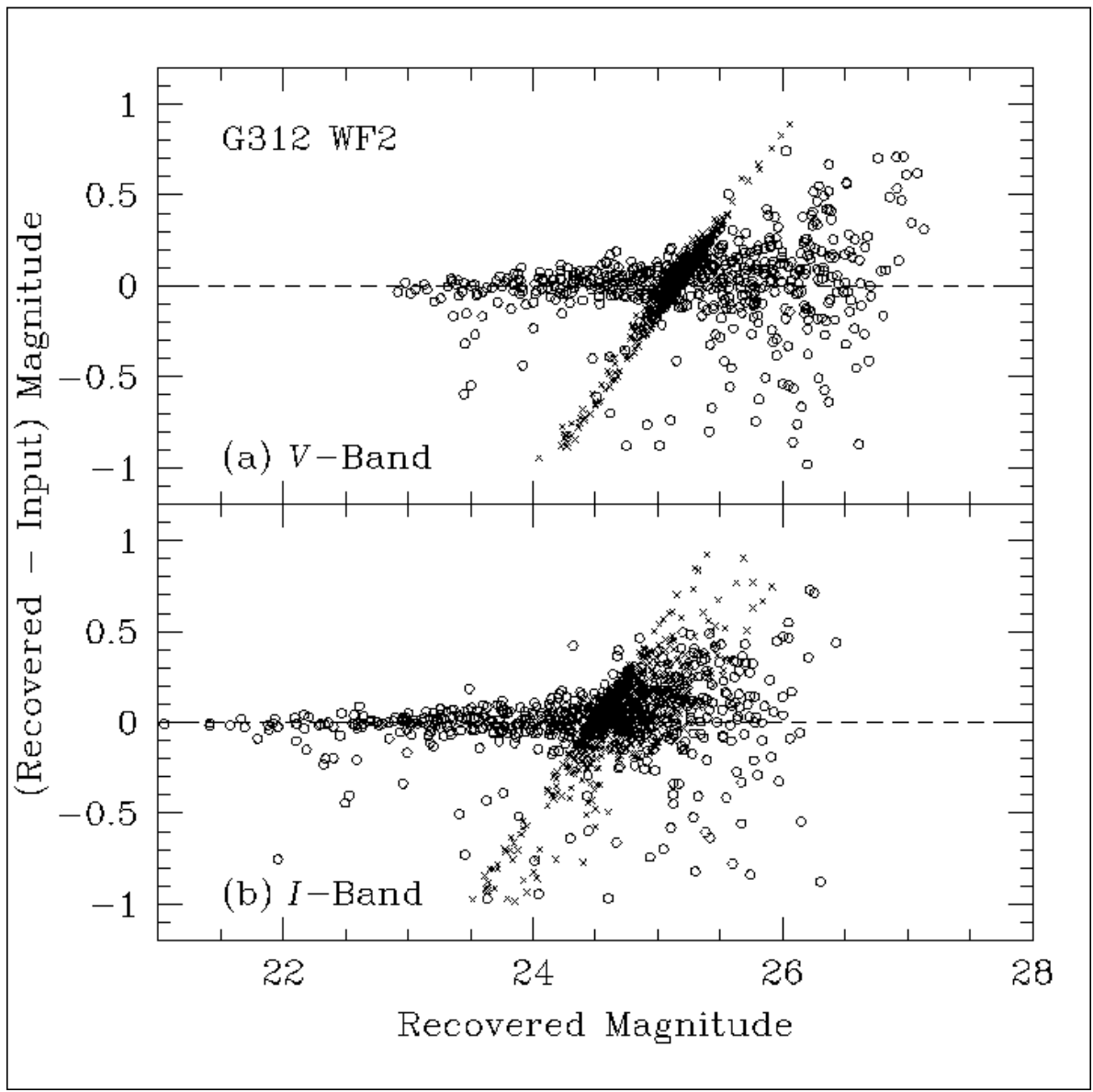}
\caption{This figure shows the scatter in the recovered magnitudes of the
artificial stars added to the WF2 background field $\sim 1\arcmin$
south of the GC G312.  The upper (a) panel shows the observed shifts
in magnitudes for the \V-band photometry while the lower (b) panel
shows the observed shifts in magnitude for the \I-band photometry.
Open circles, ``o'', denote RGB stars while crosses, ``$\times$''
denote HB stars.  Only stars with shifts of less than 1 mag are shown.
Negative magnitude shifts indicate that the star was recovered
brighter than it was input.  Positive shifts indicate the star was
recovered fainter than it was input.  KS tests show that the
distributions of recovered magnitudes for the artificial RGB and HB
stars are the same at the 98.892\% confidence level in the
\V-band and are the same at the 99.999\% confidence level in the
\I-band.  The apparent excess scatter in the HB artificial stars is
merely an artifact of the large number of artificial HB stars.
\label{FIGURE:art_shift}}
\end{figure}

	Figure~\ref{FIGURE:art_shift} shows the relationship between
the input magnitudes of the artificial stars and their recovered
magnitudes for a typical set of stars, in this case using the WF2
field near the GC G312.  Artificial stars with magnitude differences
of more than $\pm \Delta V = 1$ mag have been omitted.  This figure
plots the magnitude difference as a function of the recovered, not the
input, magnitudes of the artificial stars.  This was done because the
observed magnitudes of the real stars are the recovered magnitudes,
not the actual (or input) magnitudes.  Therefore,
Figure~\ref{FIGURE:art_shift} can be used to directly determine the
uncertainty in a star's observed magnitude.  Kolmogorov--Smirnov (KS)
tests show that the distributions of recovered magnitudes for the
artificial red-giant branch (RGB) and HB stars are the same at the
98.892\% confidence level in the \V-band and at the 99.999\%
confidence level in the
\I-band.  The apparent excess scatter in the HB artificial stars is
merely an artifact of the large number of artificial HB stars.

	Table~\ref{TABLE:art_shift_RGB} lists the systematic shifts in
$V$ and $(V-I)_0$ as well as the degree of scatter as a function of
magnitude for the artificial red-giant branch (RGB) stars of G312.
Table~\ref{TABLE:art_shift_HB} lists these quantities for the G312
artificial HB stars.  In these tables the notation $[x]$ indicates the
median value of the individual $x_i$ values.  The columns are (1) the
boundaries of the bin, (2) the median recovered magnitude, $[V]$; (3)
the median photometric uncertainty in the recovered magnitudes (from
{\sc daophot}), $[\sigma_V]$; (4) the median magnitude difference,
$[\delta_V] \equiv [V_{\rm recovered} - V_{\rm input}]$; (5) the
amount of scatter present in the artificial stars in that bin,
$\hat\sigma_V \equiv \bigl[|\delta_V - [\delta_V]|\bigr]$; columns (6)
to (9) are the same quantities for the $(V-I)_0$ color; and (10) is
the number of artificial stars in that bin.  These quantities are
identical to those described by Stetson \& Harris
\markcite{SH88}(1988) except for $\hat\sigma$, which we defined in
non-parametric terms instead of in Gaussian terms.

	Our reduction process has a slight tendency to underestimate
the brightnesses of stars near the photometric limit of our data (see
column 4 of Table~\ref{TABLE:art_shift_RGB}).  However, this
systematic shift is less than the internal scatter in the photometry
and is smaller than $\sim 0.05$ mag for RGB stars at the level of the
HB\@.  Therefore, we conclude that the systematic uncertainties in our
photometric reductions do not affect the morphology of the CMDs above
$V \sim 26$.

\begin{figure}
\plotone{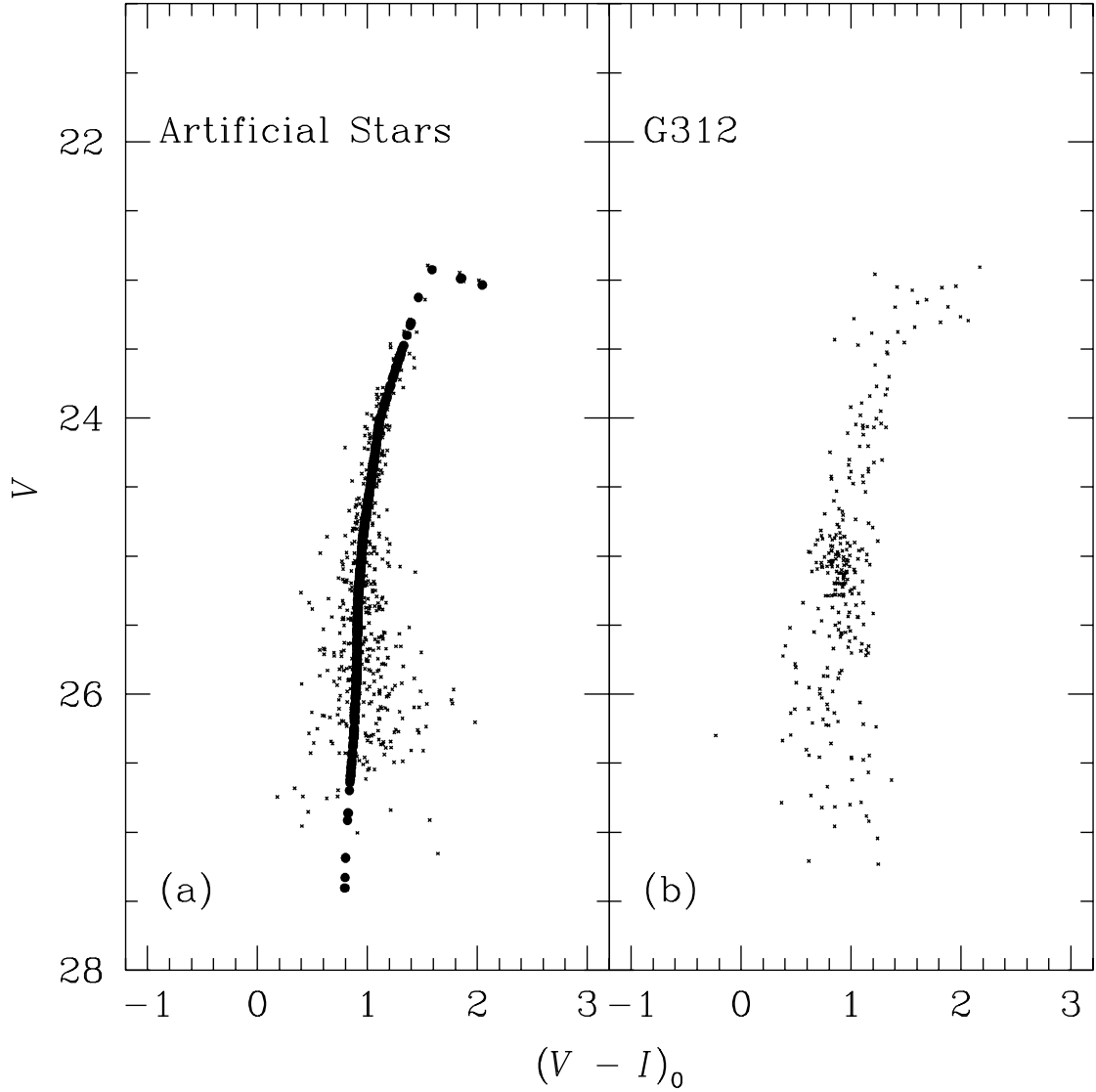}
\caption{The left-hand panel (a) shows a CMD for a set of artificial
stars added to the G312 WF3 field.  Solid circles, ``$\bullet$'',
represent the input colors and magnitudes while crosses, ``$\times$'',
represent the recovered photometry.  For clarity only RGB stars are
shown in this figure.  The right-hand panel (b) shows the observed CMD
for RGB stars in G312.  The spread in the recovered colors of the
artificial stars is approximately the same as the spread in color
observed in the G312 CMD, suggesting that the observed width of the
RGB is due to photometric uncertainties and is not intrinsic to
G312. \label{FIGURE:art_RGB}}
\end{figure}

	Figure~\ref{FIGURE:art_RGB} shows a CMD created from a set of
artificial RGB stars added to the G312 WF3 image.  The HB stars have
been omitted in order to show the increase in the scatter in the
recovered colors at fainter magnitudes.  The amount of scatter seen at
a particular magnitude is comparable to the amount of scatter seen in
the CMD of G312, which is believed to be made up of stars with a
single metallicity.  If the stellar populations in M31 GCs are
analogous to the stellar populations in Galactic GCs, then the
observed spread in G312's RGB is primarily due to photometric
uncertainties.  Figure~\ref{FIGURE:art_RGB} shows that photometric
uncertainties are sufficient to explain all the observed width of the
RGB of G312.
	
\begin{figure}
\plotone{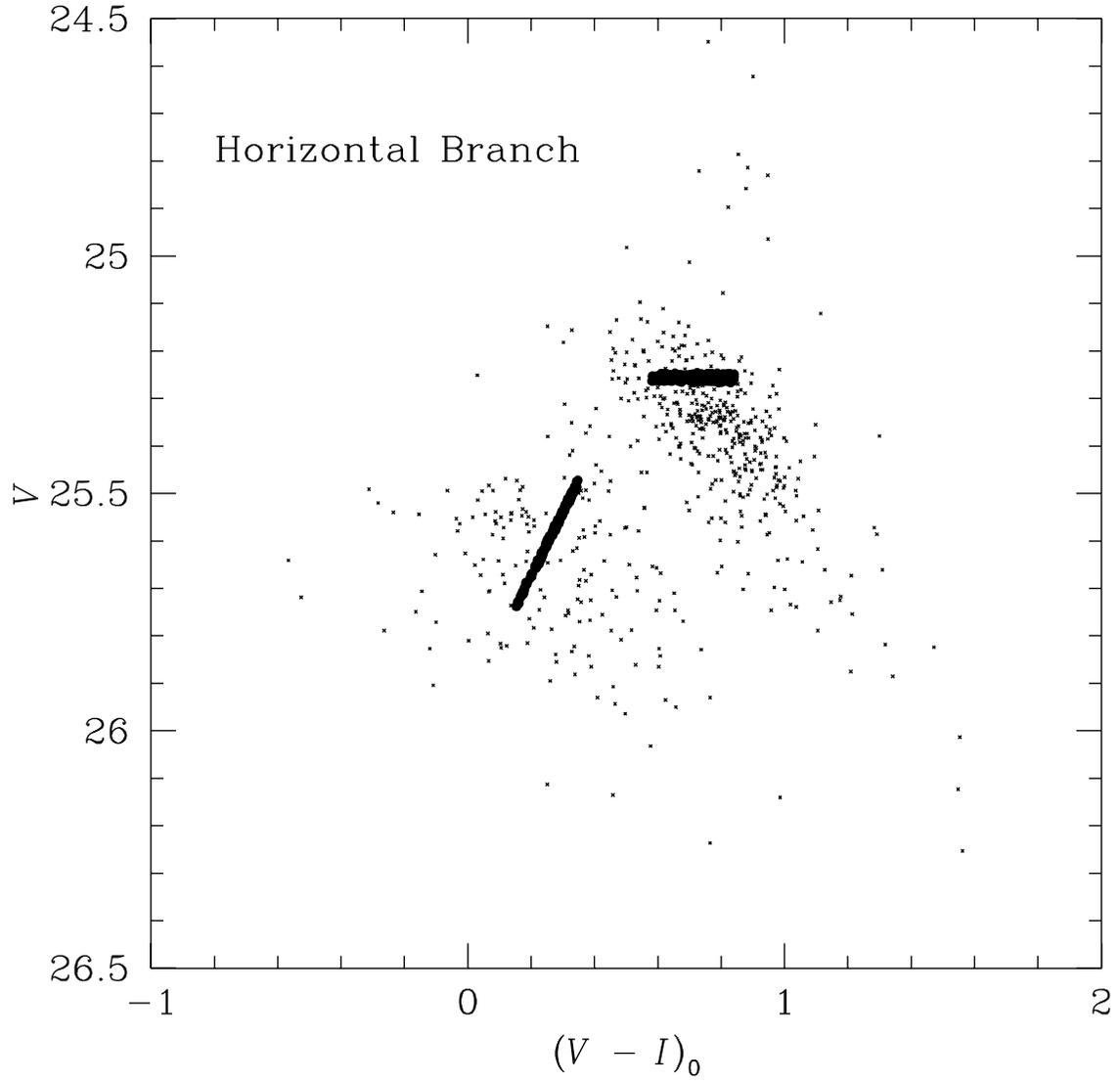}
\caption{The solid circles, ``$\bullet$'', show the input artificial
HB\@.  The HB has been divided into distinct red and blue parts to aid
in the identification of scatter from each component.  Crosses,
``$\times$'', show the recovered colors and magnitudes of the
artificial stars. \label{FIGURE:art_HB}}
\end{figure}

	Table~\ref{TABLE:art_shift_HB} and Figure~\ref{FIGURE:art_HB}
show that HB stars, because of the large color spread in the HB, can
suffer badly from systematic differences between their input and
recovered magnitudes.  Blue HB stars are scattered by as much as 0.5
mag.\ in $(V-I)_0$, while red HB stars have a tendency to be recovered
at fainter magnitudes and redder colors than they really are.  This
tends to move red HB stars into the lower half of the RGB resulting in
our star counts underestimating of the total number of red HB stars in
the GC\@.  The artificial star tests show that stars from the red and
blue sides of the HB can be scattered into the RR Lyrae gap.  This
making it difficult to distinguish RR Lyrae candidates from stars
which have simply scattered into the RR Lyrae gap.


\section{The Color--Magnitude Diagrams\label{SECTION:cmds}}

\subsection{Contamination\label{SECTION:contamination}}

\begin{figure}
\plotone{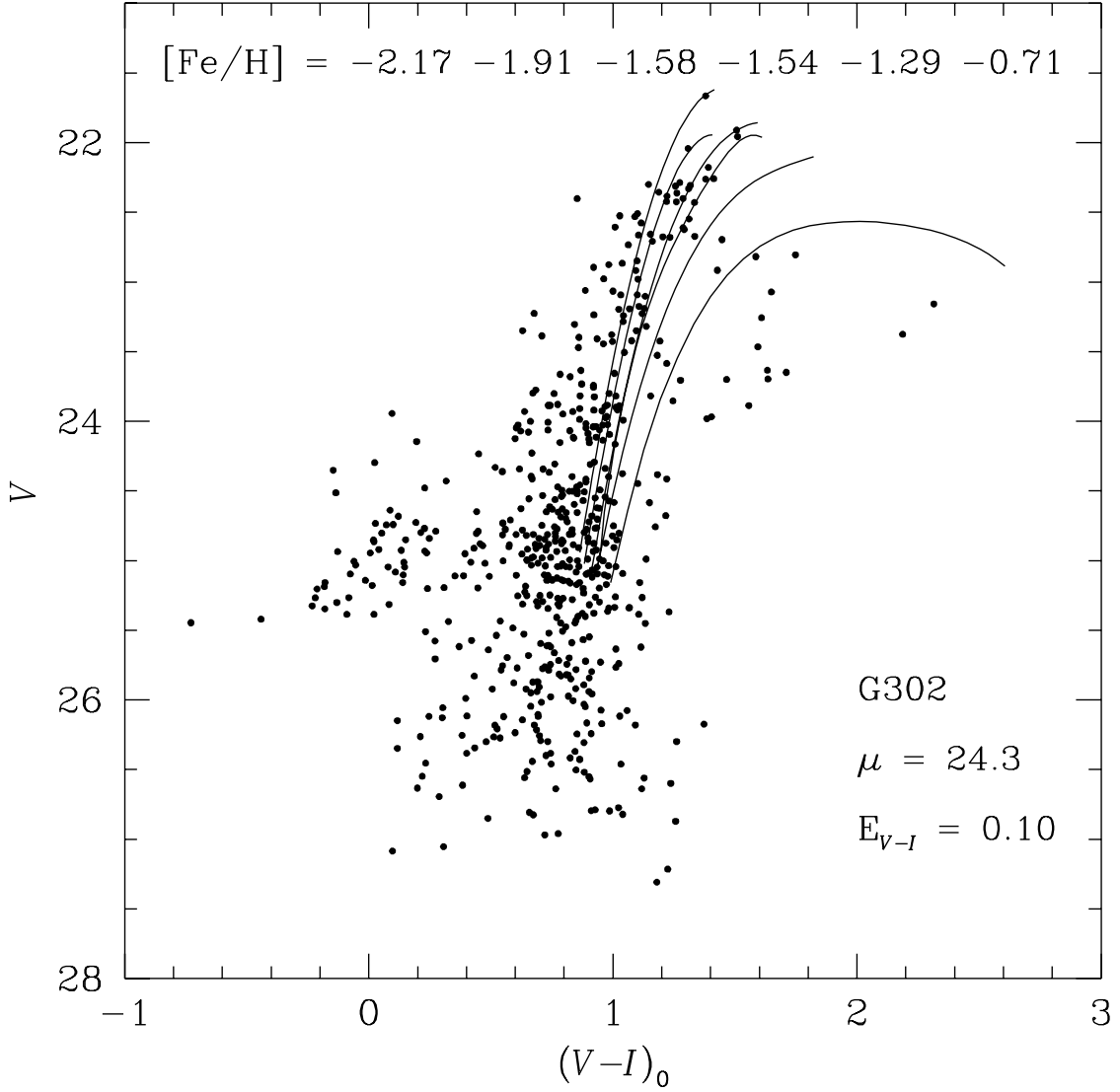}
\caption{The CMD for G302.  Only stars between 2\farcs5 $\le r \le$
$10\arcsec$ are plotted to minimize contamination from stars in the
halo of M31 and from unreliable photometry due to the extreme crowding
in the central regions of G302.  The distinct blue HB is consistent
with a metal-poor stellar population.  The RGB fiducial sequences are
of Galactic GCs and were taken from DCA\@.  The iron abundances for
each fiducial line is printed at the top of the
figure. \label{FIGURE:G302_cmd}}
\end{figure}

	There are four possible sources of contamination in the CMDs
of the GCs.  The first three---Galactic halo stars, background
galaxies, and M31 disk stars---are discussed in \sect 3.1 of HFR\@.
By scaling those results to the area on the sky covered by each GC we
estimate that at most $15 \pm 4$ of the objects with $V \le 24$ in
each of
Figures~\ref{FIGURE:G302_cmd}~(G302)~and~\ref{FIGURE:G312_cmd}~(G312)
are due to contamination from these three sources.

	The fourth source of contamination is the halo of M31 itself.
To estimate the number of halo stars that should be visible in the CMD
for each GC we computed the stellar surface densities in the WF2 and
WF4 CCDs for each field.  These CCDs were centered $\sim 75\arcsec$
from the center of the WF3 CCD (and thus the GCs).  Since G302 and
G312 have tidal radii of $\sim 10\arcsec$ the WF2 and WF4 CCDs will
not suffer from contamination from GC stars.  The mean background for
M31 halo stars with $V \le 24$ near G302 is $\Sigma_{\rm bkgd} =
0.0654 \pm 0.0024$ stars/\sq\arcsec, corresponding to $20 \pm 5$ halo
stars in the RGB for G302.  Near G312 the background is $\Sigma_{\rm
bkgd} = 0.0174 \pm 0.0012$ stars/\sq\arcsec, corresponding to $5 \pm
1$ stars in the RGB of G312.  These estimates are reasonably
consistent with the numbers of objects in the upper halves of
Figures~\ref{FIGURE:G302_cmd}~and~\ref{FIGURE:G312_cmd} that do not
fall on either the RGB or the asymptotic-giant branch (AGB) of each
GC\@.

	On the basis of these calculations we expect that
contaminating objects make up $\lesssim 7$\% of the objects in the
G302 CMD\@.  Similarly contaminating objects make up $\lesssim 16$\%
of the objects in the G312 CMD\@.  These numbers are consistent with
the number of discrepant objects in each CMD so we conclude that the
morphologies of the CMDs in
Figures~\ref{FIGURE:G302_cmd}~and~\ref{FIGURE:G312_cmd} are real and
not due to contamination.

\subsection{G302\label{SECTION:cmds_G302}}

	Figure~\ref{FIGURE:G302_cmd} shows the CMD for the GC G302.
Stars located between $2\farcs5 \le r \le 10\arcsec$ from the center
of the cluster are shown.  The $r = 10\arcsec$ cut-off approximately
corresponds to the tidal radius of the GC while the $r = 2\farcs5$
cut-off is the point where extreme crowding conditions make it
impossible to obtain reliable stellar photometry.  To estimate the
metallicity of G302 we took the dereddened fiducial RGB sequences for
several Galactic GCs (from Da Costa \& Armandroff \markcite{DA90}1990,
hereafter referred to as DCA), shifted them to the distance of M31
($\mu_0 = 24.3$), and overlaid them onto Figure~\ref{FIGURE:G302_cmd}.
The lack of a turn-over at the top of the G302 RGB is consistent with
a metal-poor population with an iron abundance of $\FeH \simeq -1.5$
to $-2.2$.  We interpolated between the DCA fiducial sequences to
obtain $\FeH = -1.85 \pm 0.12$ for G302, where the uncertainty is
one-third of the spread in the RGB and corresponds to approximately
one standard deviation if we assume the spread in the RGB is due to
normally distributed errors in the photometry.  Our estimate of the
iron abundance for G302 agrees with spectroscopic estimates by Huchra
\etal \markcite{HB91}(1991, hereafter referred to as HBK), $\FeH =
-1.76 \pm 0.18$, and de Freitas Pacheco \markcite{dFP97}(1997), $\FeH
= -1.80 \pm 0.2$.

	 AGB stars in G302 can be seen to the blue of the
low-metallicity fiducial sequences in Figure~\ref{FIGURE:G302_cmd}.
Unfortunately the AGB and the RGB blend together so it is not possible
to unambiguously distinguish AGB stars from RGB stars.  This is due to
photometric scatter ($\sigma_{V-I} \simeq 0.06$ at $V = 23$) and
contamination from RGB stars in the halo of M31.

	No stars brighter than the tip of the RGB are present in the
data.  The lack of such super-luminous stars supports the hypothesis
that GCs in M31 have ages similar to those of the Galactic GCs.
Spectral line blanketing in cool giants can cause very red giants to
appear fainter than the bluer, less evolved, RGB stars.  This can
result in some massive evolved stars appearing to have $(V-I)$ colors
considerably redder than the rest of the GC stars.  Our artificial
star tests suggest that we are able to detect any stars above the HB
so we are confident that we have not missed any super-luminous AGB
stars in the outer regions of G302.  However, since these stars, being
more evolved, will have slightly more massive progenitors than the
rest of the RGB and AGB stars, it is possible that mass segregation
will have concentrated them in the inner $r = 2\farcs5$ of the GC
making them undetectable with our data.  To date no super-luminous AGB
stars have been detected in any M31 GC (\eg see Rich \etal
\markcite{RM96}1996, Fusi Pecci \etal \markcite{FP96}1996, Jablonka
\etal \markcite{JB97}1997 for CMDs obtained using {\sl HST\/} observations).
This strongly supports the idea that M31 GCs have comparable ages to
those of Galactic GCs.  We wish to emphasize, however, that although
we see no evidence in our data for G302 having an age that is any
different from the ages of the Galactic GCs, we can not rule out the
presence of super-luminous stars in the inner regions of G302.

	Further evidence that G302 is an old GC with an age similar to
that of the Galactic GCs comes from de Freitas Pacheco
\markcite{dPF97}(1997).  That study used integrated spectra and
single-population stellar populations models to estimate the ages of
twelve M31 GCs---including G302.  The result was a mean age of $15 \pm
2.8$ Gyr for their sample of twelve M31 GCs and an age of $17.0 \pm
2.9$ Gyr for G302.

	The weighted mean magnitude of the stars in the RR Lyrae gap
is $\overline{V_{\rm RR}} = 24.93 \pm 0.09$ (standard deviation).
Using this, our derived iron abundance of $\FeH = -1.85$, and

\begin{equation}
M_V({\rm RR}) = 0.20 \FeH + 0.98
\label{EQUATION:FeH_MV_CD96}
\end{equation}

\noindent
(Chaboyer \etal \markcite{CD96}1996), we get a distance modulus of
$\mu_0 = 24.32 \pm 0.09$ for G302.

\begin{figure}
\plotone{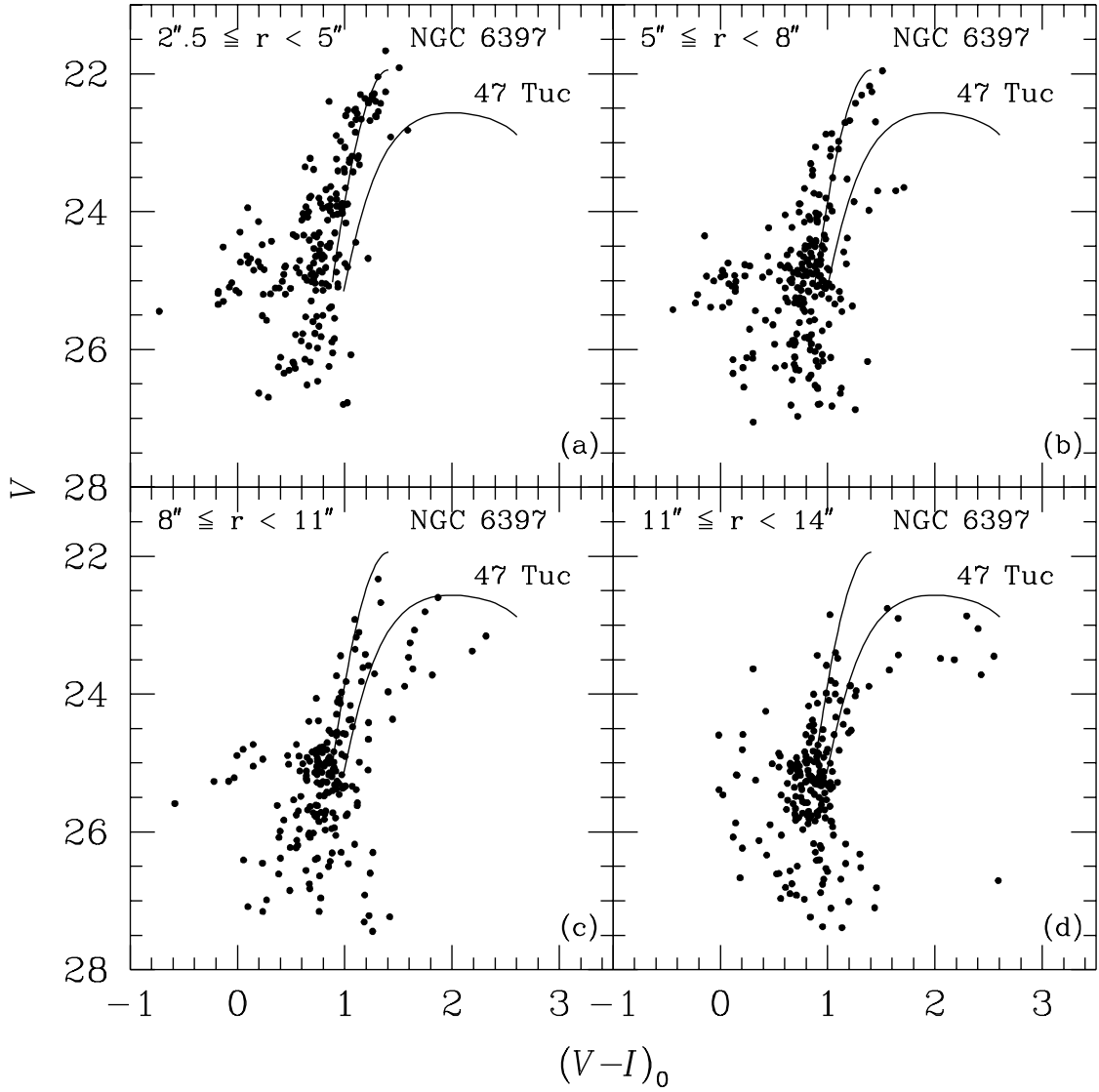}
\caption{This figure shows CMDs for four annuli centered on G302.
The fiducial sequences are from DCA\@. \label{FIGURE:G302_cmd_rings}}
\end{figure}

\begin{figure}
\plotone{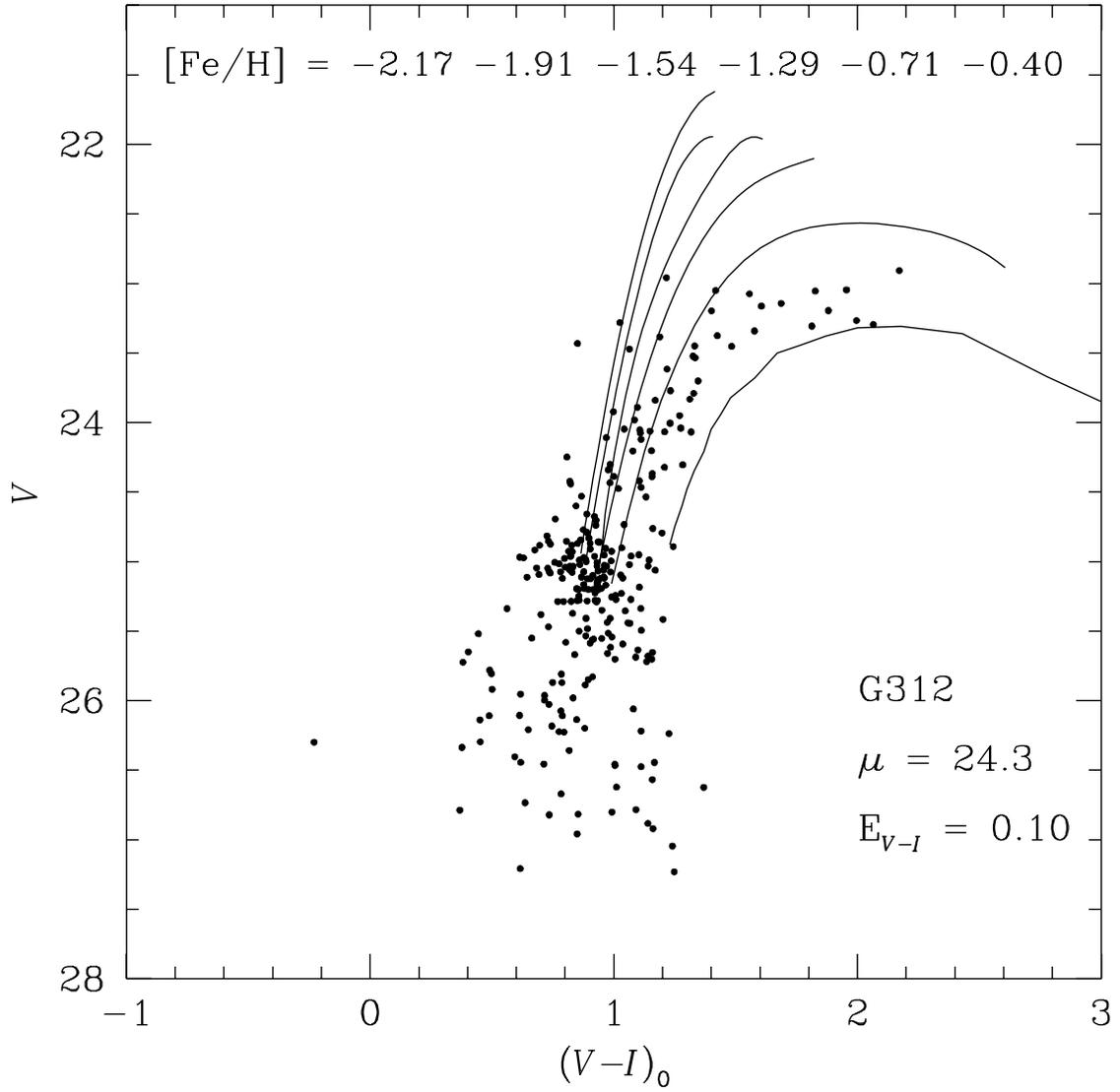}
\caption{The CMD for G312.  As in Figure~\ref{FIGURE:G302_cmd} only stars
between $2\farcs5 \le r \le 10\arcsec$ are plotted.  The fiducial
sequences are from DCA with the exception of the $\FeH = -0.40$
fiducial which is from Bergbusch \& VandenBerg
\protect\markcite{BV92}(1992).  The lack of a blue HB is consistent
with G312 being metal-rich. \label{FIGURE:G312_cmd}}
\end{figure}

	In order to identify the outer limiting radius of G302, and to
search for any radial dependence in the morphology of the various
branches of the CMD, we constructed CMDs in four annuli centered on
the cluster.  These CMDs are shown in
Figure~\ref{FIGURE:G302_cmd_rings}.  The fiducial sequences plotted in
Figure~\ref{FIGURE:G302_cmd_rings} are those of NGC 6397 ($\FeH =
-1.91$) and 47 Tuc ($\FeH = -0.71$) taken from DCA\@.  For $r \lesssim
8\arcsec$ the NGC 6397 fiducial provides a good fit to the G302 RGB\@.
The GC appears to end at a radius somewhere between $8\arcsec$ and
$11\arcsec$, in agreement with what is found by fitting MK models (see
\sect~\ref{SECTION:structure}).  Beyond $11\arcsec$ the RGB of the metal-rich
halo population is clearly visible.  The apparent increase in the
number of blue HB stars in the inner regions of G302 is an illusion.
A KS test shows that the radial distribution of blue HB stars is the
same as that of the RGB stars at the 99.91\% confidence level.

\subsection{G312\label{SECTION:cmds_G312}}

	Figure~\ref{FIGURE:G312_cmd} shows the CMD for the GC G312.
As with G302 only stars between $2\farcs5 \le r \le 10\arcsec$ are
shown since star counts and the surface brightness profiles suggest
that the MK tidal radius of G312 occurs at $\sim 10\arcsec$ (Holland
\etal \markcite{HF93}1993, Holland \markcite{H97}1997).  Interpolating
between the fiducial giant branches suggests that G312 has an iron
abundance of $\FeH = -0.56 \pm 0.03$, comparable to the mean iron
abundance of stars in the halo of M31.  This is an agreement with the
spectroscopic iron abundance determined by HBK of $\FeH = -0.70 \pm
0.35$.  The quoted uncertainty in our iron abundance estimate is the
formal uncertainty in the interpolation based on the observed width of
the RGB\@.  It does not include uncertainties in the iron abundances
of the fiducial RGB sequences or uncertainties in the distance to
G312.  In light of this the quoted uncertainty should be considered a
lower limit on the true uncertainty in our estimate of $\FeH$.  The
small numbers of RGB stars falling under the metal-poor fiducial
sequences are consistent with the expected degree of contamination
(see
\sect~\ref{SECTION:contamination}) from disk and metal-poor halo stars
in M31 (see \sect 3.2 of HFR for details on the metallicity
distribution of halo stars in M31).

	The HB of G312 is a red clump located slightly to the blue of
the RGB\@.  No HB stars bluer than $(V-I)_0 = 0.6$ are present so it
is not possible to determine the magnitude of the HB in the RR Lyrae
gap.  Therefore we estimated $\overline{V_{\rm RR}}$, by taking the
weighted mean magnitude of the red HB stars, to be $\overline{V_{\rm
HB}} = 25.09 \pm 0.09$.  In order to get $\overline{V_{\rm RR}}$ we
added a correction to $\overline{V_{\rm HB}}$ (see \eg Lee \etal
\markcite{LD94}1994) of $\Delta V^{\rm rich}_{\rm HB} = 0.08$ (see
Sarajedini \etal \markcite{SD95}1995, Ajhar \etal
\markcite{AG96}1996).  This gives $\overline{V_{\rm RR}} = 25.17 \pm
0.09$ for G312.  Using the iron abundance we derived from the shape of
the RGB the Chaboyer \etal \markcite{CD96}(1996) relation
(Eq.~\ref{EQUATION:FeH_MV_CD96}) gives a distance modulus of $\mu_0 =
24.30 \pm 0.09$.

	Since so few halo stars are expected in the upper RGB of G312,
the observed width of the RGB provides a good estimate of the true
photometric uncertainties in our observations.  Photometric and
spectroscopic evidence strongly suggests that all stars in a single
Galactic GC have the same metallicity and there is no evidence to
suggest that this should be different for GCs in M31.  The mean width
of the RGB is $\sigma_{V-I} \sim 0.08$, consistent with what is
predicted by the artificial star simulations (see
\sect~\ref{SECTION:art_stars}).  Therefore, we believe that the RGB
width in Figure~\ref{FIGURE:G312_cmd} is typical of a
single-metallicity population for our WF observations.  This Figure
can be compared to Figures 4 and 5 of HFR to show that the large range
of metallicities seen in the halo of M31 is not merely an artifact of
photometric uncertainties along the halo RGB\@.

\begin{figure}
\plotone{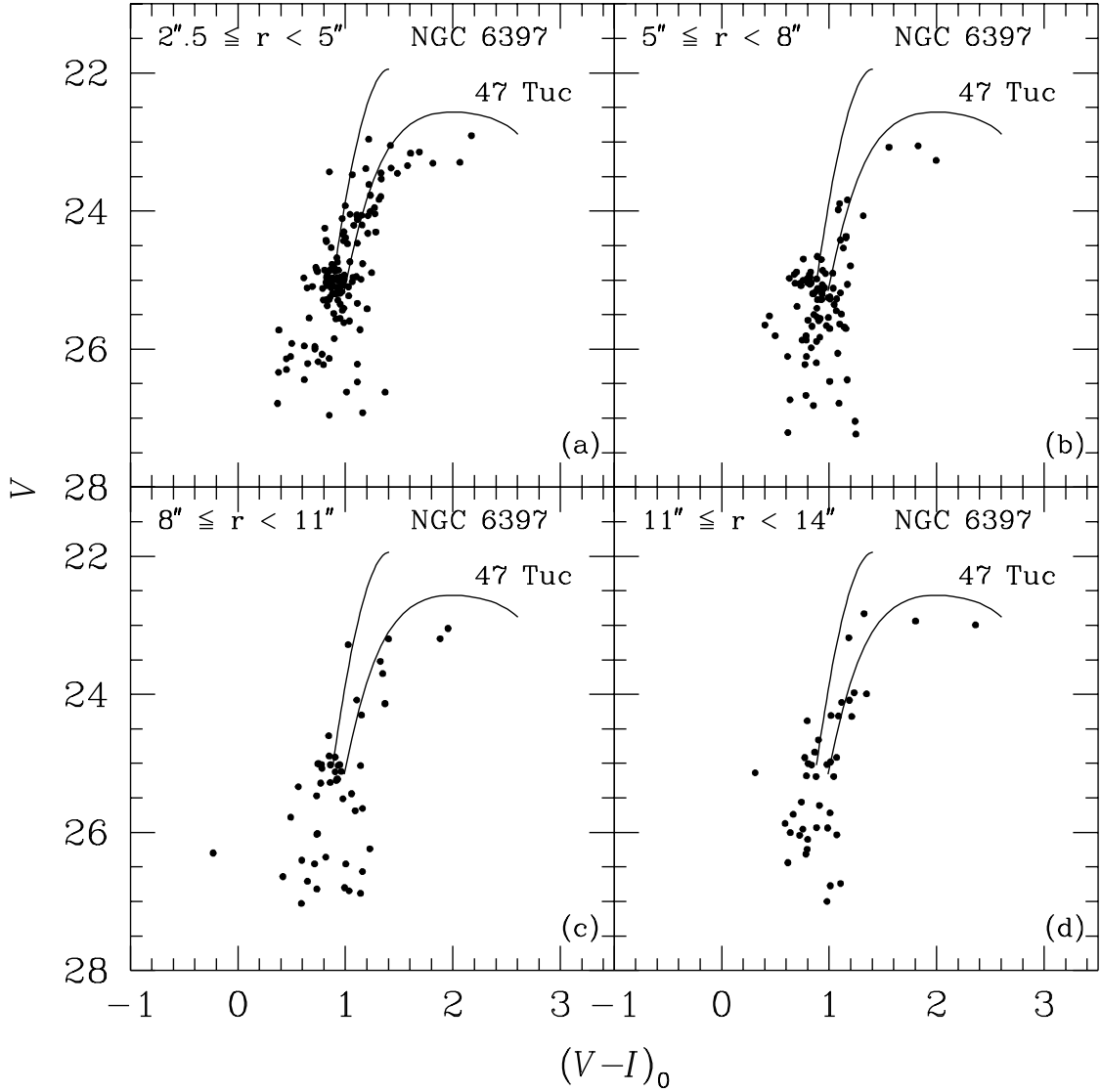}
\caption{This figure shows CMDs for four annuli centered on G312.
Since G312 has a metallicity similar to the mean metallicity of the
M31 halo stars in the line of sight near G312 it is not possible to
determine the limiting radius of G312 in this figure.  The fiducial
sequences are from DCA\@. \label{FIGURE:G312_cmd_rings}}
\end{figure}

	Figure~\ref{FIGURE:G312_cmd_rings} shows no radial dependence
in the morphology of the G312 CMD\@.

\subsection{The Color--Iron Abundance Relation\label{SECTION:vi-feh}}

\begin{figure}
\plotone{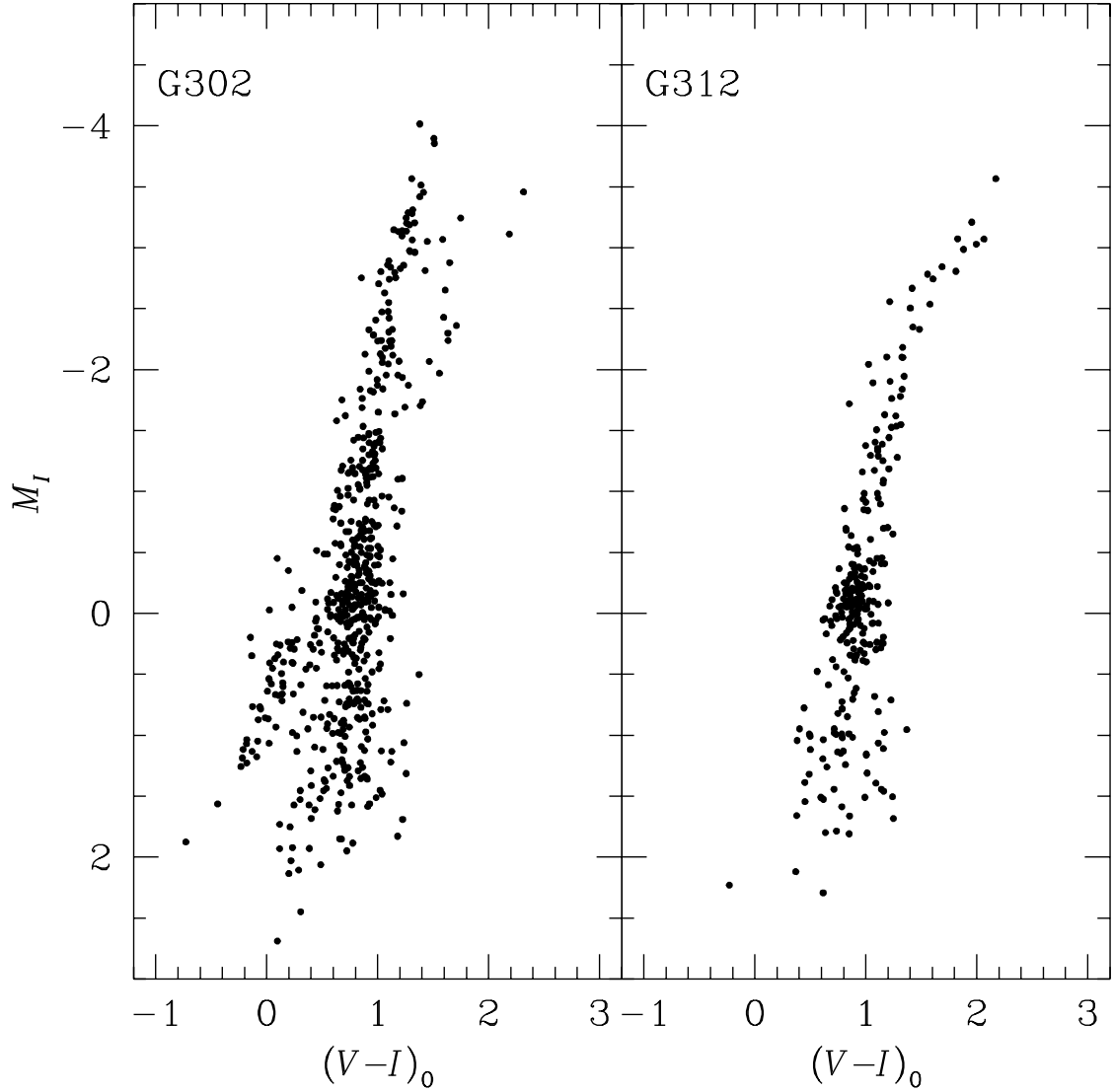}
\caption{This figure shows the $(M_I,(V-I)_0)$ CMDs for G302 and
G312.  A distance modulus of $\mu_0 = 24.3$, a reddening of $E_{V-I} =
0.1$, and an extinction $A_I = 0.19$ was assumed as discussed in
\sect~\ref{SECTION:data}. \label{FIGURE:ivi_cmds}}
\end{figure}

\begin{figure}
\plotone{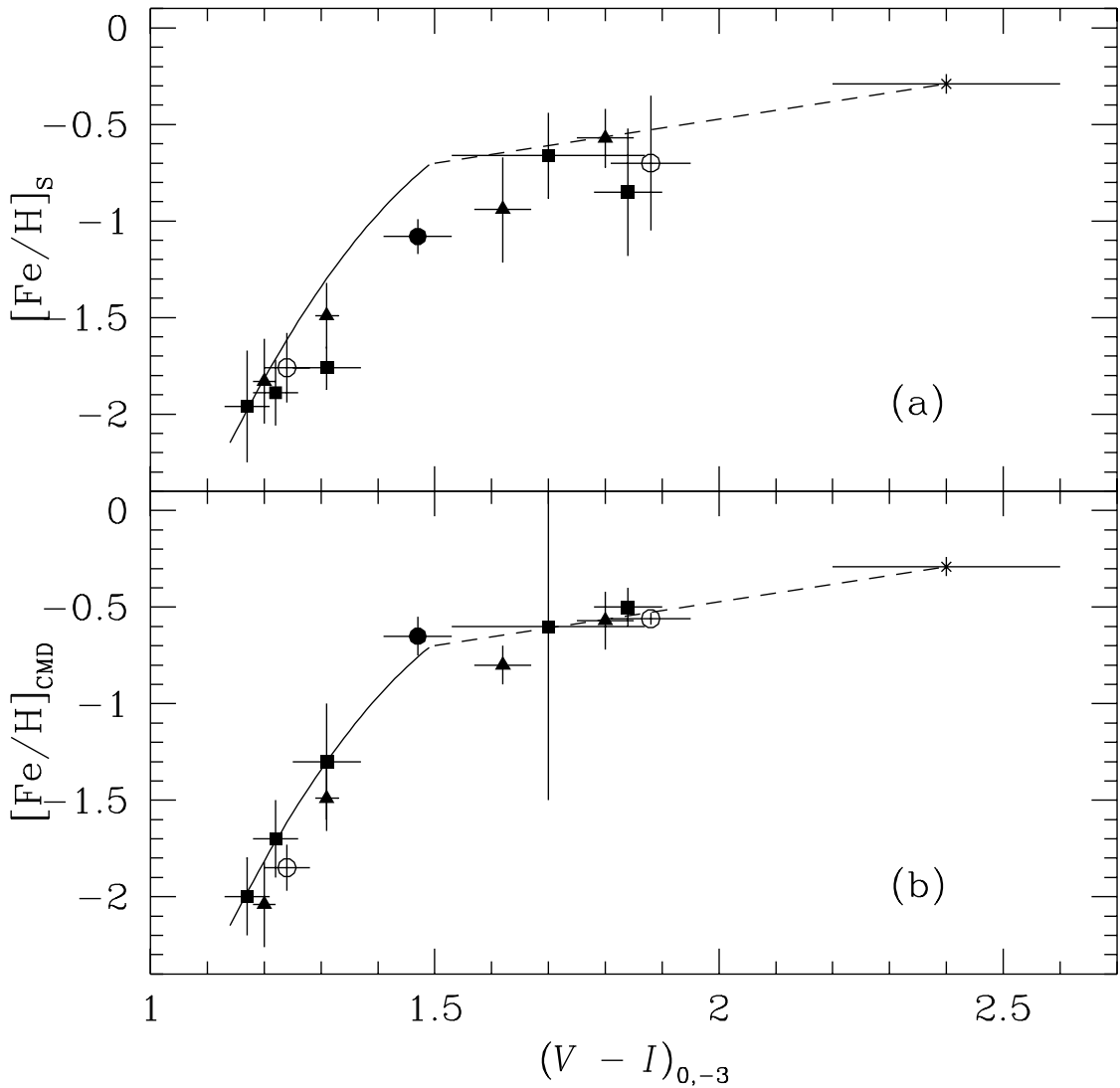}
\caption{The open circles represent the data for G302 and G312
from this paper.  The solid circle is data from Rich \etal
\protect\markcite{RM96}(1996), the solid squares are data from Couture
\etal \protect\markcite{CR95}(1995), and the solid triangles are data from
Ajhar \etal \protect\markcite{AG96}(1996).  The horizontal error bars
only include the uncertainty due to the width of the RGB, not any
uncertainties in the reddening or distance modulus of the individual
GCs.  The solid line shows the DCA relationship between iron abundance
and the $(V-I)_{0,-3}$ color while the dashed line shows our extension
to this relationship to include the metal-rich Galactic GC NGC 6553
(indicated with a ``*'').  Spectroscopic $\FeH$ values from HBK are
plotted on the ordinate of panel (a) and $\FeH$ values derived from
the CMDs are plotted on the ordinate of panel (b).  Notice that in (b)
all the GCs fall within 1-$\sigma$ of the extended DCA relation except
for G108, which is $\sim 3$-$\sigma$ redder than predicted by DCA\@.
\label{FIGURE:vi-feh}}
\end{figure}

	Figure~\ref{FIGURE:ivi_cmds} shows the CMDs for G302 and G312
after converting the calibrated magnitudes to absolute \I-band
magnitudes.  This was done to determine the color of the RGB for each
GC at $M_I = -3$, hereafter referred to as $(V-I)_{0,-3}$.  DCA found
a strong relationship between iron abundance and $(V-I)_{0,-3}$ for
Galactic GCs with $-2.2 \le \FeH \le -0.7$.
Figure~\ref{FIGURE:vi-feh}a shows the relationship between
$(V-I)_{0,-3}$ and $\FeH_{\rm S}$ for G302 and G312 and several other
M31 GCs with published CMDs.  We have adjusted the published data to
the distance modulus used in this paper ($\mu_0 = 24.3$).  The
$\FeH_{\rm S}$ values are the spectroscopically determined iron
abundances of HBK, so the $(V-I)_{0,-3}$ color and iron abundance are
determined completely independently of each other for each GC\@.
Color, iron abundance, and reddening data for each GC is listed in
Table~\ref{TABLE:vi-feh} along with the source of the CMD; the
projected distance of each GC from the center of M31; and $Y$, the
projected distance of each GC from the major axis of M31 with positive
$Y$ being on the northwest side of the major axis of M31.

	Since several of the M31 GCs in Figure~\ref{FIGURE:vi-feh}a
have iron abundances at least as high as that of 47 Tuc we extended
the DCA relationship to higher iron abundances.  Extending this
relationship to GCs more metal-rich than 47 Tuc ($\FeH = -0.71$) is
somewhat difficult since the RGB in the $((V-I)_0,M_I)$ plane becomes
asymptotically flat as the iron abundance increases.  This makes it
difficult to define $(V-I)_{0.-3}$ since, as $\FeH$ approaches the
Solar value, this point occurs in the horizontal portion of the RGB\@.
In order to extend the DCA relationship to higher iron abundances we
determined $(V-I)_{0,-3}$ for the metal-rich Galactic GC NGC 6553
using the $(I,(V-I))$ CMD of Ortolani \etal \markcite{OB90}(1990).  We
assumed a distance modulus of $\mu_0 = 13.35$ (Guarnieri \etal
\markcite{GM95}1995), a reddening of $E_{B-V} = 0.78$ (Bico \& Alloin
\markcite{BA86}1986), and an iron abundance of $\FeH = -0.29$ (Zinn \&
West \markcite{ZW84}1984).  Using this data we found $(V-I)_{0,-3} =
2.4 \pm 0.2$ for NGC 6553.  This point lies just below the region
where the RGB becomes horizontal so there is a significant uncertainty
in the value of $(V-I)_{0,-3}$ for NGC 6553 solely due to the width of
the RGB in $(V-I)_0$ at $M_I = -3$.  We extrapolated the DCA relation
to $\FeH = -0.29$ simply by connecting the $\FeH = -0.71$ end of the
DCA relation to the location of NGC 6552.  The resulting relation is

\begin{equation}
\FeH = -15.16 + 17.0(V-I)_{0,-3} - 4.9(V-I)_{0,-3}^2
\end{equation}

\noindent
for $-2.2 \le \FeH \le -0.7$ (from DCA) and

\begin{equation}
\FeH =  -1.36 + 0.44(V-I)_{0,-3}
\end{equation}

\noindent
for $-0.7 \le \FeH \le -0.29$ (our extension).  We wish to stress that
our extension to high iron abundances is an estimate and is intended
only to provide a reasonable estimate of what happens to the DCA
relation at high iron abundances.

	If the spectroscopic iron abundances are adopted then nine of
the twelve M31 GCs are redder than predicted by the extended DCA
relation.  This discrepancy is present at all iron abundances but is
most noticeable for the metal-rich GCs.  The individual uncertainties
in $\FeH_{\rm S}$ for all the GCs except G1 and G327 are large enough
to account for the difference between their locations in
Figure~\ref{FIGURE:vi-feh}a and the predictions of the extended DCA
relation.  A recent paper by de Freitas Pacheco
\markcite{dfP97}(1997) gives iron abundances for twelve GCs in the
M31 system.  Most of their iron abundances are higher than the HBK
values by $\sim 0.2$ dex.  An examination of the CMDs for each GC
shows that several GCs have RGBs that are significantly redder and
flatter than would be expected from their spectroscopic iron
abundances.  This is particularly noticeable for the more metal-rich
GCs.  To test whether errors in the spectroscopically determined iron
abundances are sufficient to account for the discrepancies between the
GCs and the extended DCA relation in Figure~\ref{FIGURE:vi-feh}a, we
replotted the GCs using iron abundances determined from the shapes of
the RGBs.  This data is shown in Figure~\ref{FIGURE:vi-feh}b.  All GCs
now fall within their uncertainties of the extended DCA relationship
with the exception of G108, which has a $(V-I)_{0,-3}$ value nearly
$3$-$\sigma$ redder than its predicted value.  Alternately, the iron
abundance Ajhar \etal \markcite{AG96}(1996) derived from G108's CMD
could be in error.  However, their Figure 20 shows the upper portion
RGB of G108 to be not quite as flat than that of 47 Tuc, indicating
that G108 has slightly less iron than 47 Tuc does.  It would be
difficult, using their data, to fit a fiducial to the G108 RGB that is
flatter (and thus more iron rich) than the 47 Tuc fiducial.
Therefore, we believe that the estimated iron abundance of G108 is
correct.

	One possible explanation for the red color of G108 is that the
\ion{H}{1} in M31 extends $\sim 15$ kpc beyond the outer edge of the
optical disk, which is past the location of G108.  Cuillandre \etal
\markcite{CL97}(1997) found dust beyond the optical limit of the M31
disk which correlates with the distribution of \ion{H}{1} beyond the
disk.  Since G108 is located within the extended disk of \ion{H}{1}
around M31 it is possible that G108's $(V-I)_{0,-3}$ color may be due
to internal reddening in the extended disk of M31.  Using the
\ion{H}{1} maps of Emerson \markcite{E74}(1974), and the relationships
between \ion{H}{1} column density and reddening given in Cuillandre
\etal \markcite{CL97}(1997), we find an excess reddening of $E_{V-I} =
0.14 \pm 0.02$.  However, an examination of the CMD for G108 (Figure
20 of Ajhar \etal \markcite{AG96}1996) suggests the internal reddening
from M31 does not exceed $E_{V-I} \sim 0.05$.  This discrepancy may be
due to the patchy nature of the \ion{H}{1} distribution in the outer
regions of M31's extended disk.  Brinks \& Bajaja
\markcite{BB86}(1986) found many small regions in the \ion{H}{1}
distribution in M31 where the \ion{H}{1} column densities were
significantly less than the surroundings.  These holes were typically
a few arcminutes in diameter, smaller than the beam-size used to
construct the Emerson \markcite{E74}(1974) \ion{H}{1} maps.  It is
possible that G108 (which has an angular diameter of $\sim 0\farcm4$)
is located in a hole in the \ion{H}{1} distribution that is smaller
than the resolution of the \ion{H}{1} maps.  If this is the case, the
Emerson \markcite{E74}(1974) \ion{H}{1} maps will lead to our
overestimating the reddening toward G108 and our estimate of the
excess reddening towards G108 would not be inconsistent with the
published CMD for that GC\@.


\section{Luminosity Functions\label{SECTION:lfs}}

\begin{figure}
\plotone{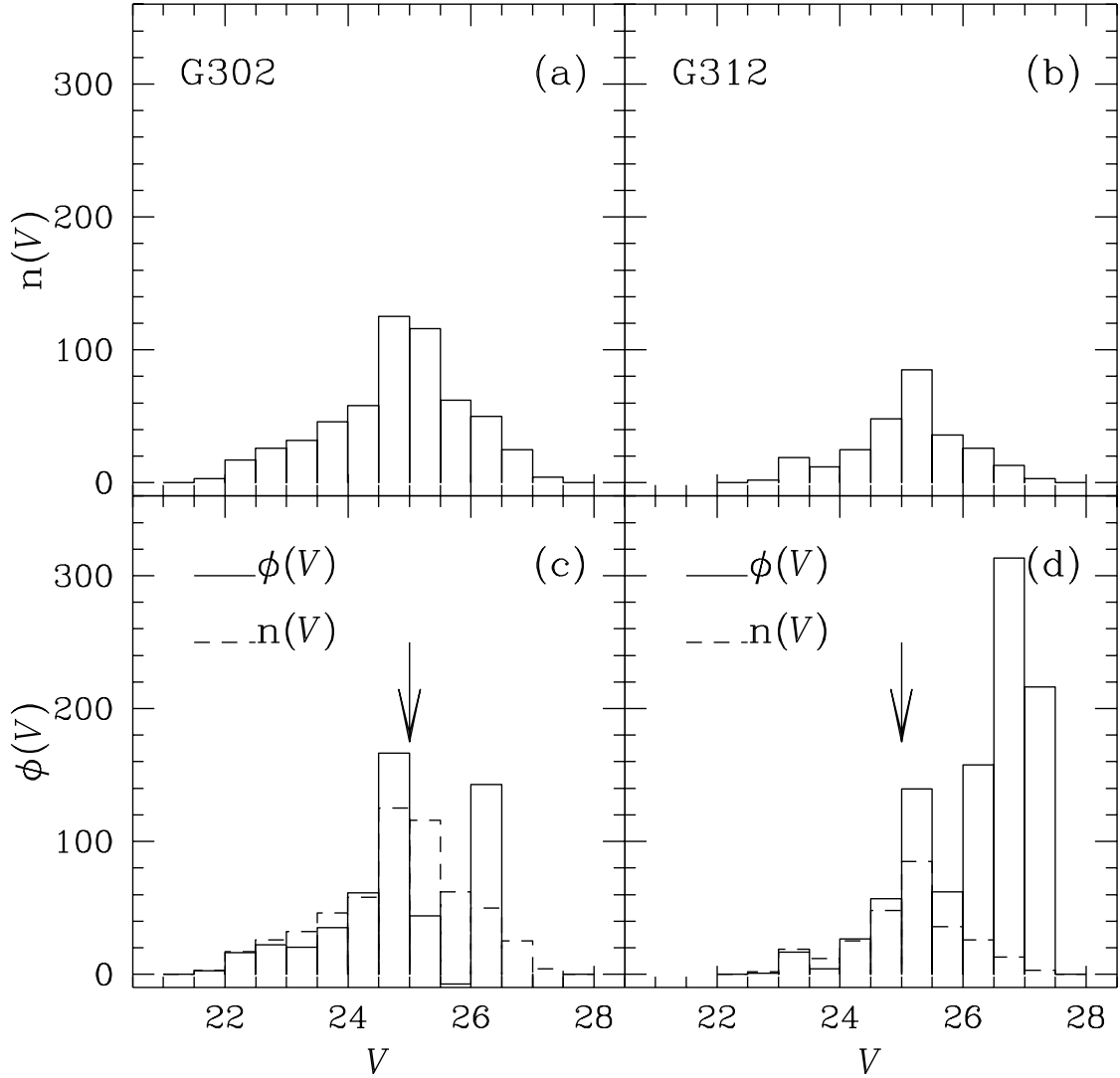}
\caption{The upper panels, (a) and (b), show the observed LFs for
G302 and G312 respectively.  Only stars between $2\farcs5
\le r < 12\farcs5$ were used to construct these LFs.  The lower
panels, (c) and (d) show the completeness-corrected LFs with the
completeness-corrected background LF subtracted (solid lines) and the
observed LFs (dashed lines).  The completeness-corrected LFs become
unreliable below the level of the HB (indicated with an
arrow). \label{FIGURE:lfs}}
\end{figure}

	The raw, $n(V)$, and completeness-corrected, $\phi(V)$,
$V$-band stellar LFs for G302 and G312 are presented in
Table~\ref{TABLE:lfs}.  The raw LFs include all stars within $2\farcs5
\le r < 12\farcs5$ of the center of each GC\@.  The background LF for
each GC was obtained by taking the LF for the M31 halo near each GC
(from HFR) and scaling them to the area covered by each GC\@.  The
scaled background LF was subtracted from the raw LF before the effects
of incompleteness were considered.  Subtracting the raw background LF
assumes that the completeness corrections in the WF2 and WF4 fields
are the same as in the WF3 fields.  An examination of
Figure~\ref{FIGURE:lfs} shows that the completeness corrections are
small for stars with $V \lesssim 25$.  Therefore the raw background
luminosity function instead of the completeness-corrected background
luminosity function will not alter the results.  The raw and
completeness-corrected LFs are related by $n(V) = {\bf P} \phi(V)$
where ${\bf P}$ is the finding-probability matrix.  ${\bf P}$
describes the probability of a star with a true magnitude of $V_{\rm
in}$ being recovered with a magnitude of $V_{\rm rec}$.

	Our artificial star tests suggest that completeness
corrections need to be applied to the raw GC and background LFs.
These corrections are small for $V \lesssim 25$, but the artificial
star tests suggest that bin-jumping becomes significant at
approximately the level of the HB and needs to be compensated for in
order to accurately reconstruct the LF\@.

	Given ${\bf P}$ there is no ideal method for reconstructing a
stellar LF from incomplete star counts.  For a review of the problem
see Fahlman \markcite{F93}(1993).  Ideally the true LF, $\phi(V)$, can
be obtained from the observed LF, $n(V)$, using $\phi(V) = {\bf
P}^{-1}n(V)$ as described in Drukier \etal \markcite{DF88}(1988) and
Mighell \markcite{M90}(1990).  Unfortunately, many faint artificial
stars are lost in the noise and thus not recovered at all.  This
results in some artificial stars bin-jumping off the faint end of the
finding-probability matrix so there is insufficient information to
reliably invert ${\bf P}$.  An alternative to computing ${\bf P}^{-1}$
is to use a model LF, $\phi^{\prime}(V)$, and solve the forward
equation $n^{\prime}(V) = {\bf P}\phi^{\prime}(V)$.  The estimated LF
can then be adjusted until $\Delta n \equiv n - n^{\prime}$ is
minimized.  This approach will not accurately reconstruct the faint
end of the LF since ${\bf P}$ does not contain sufficient information
at faint magnitudes, due to bin-jumping by faint stars, to do this.
However, unlike the inversion method, it will work if ${\bf P}$ is
nearly singular and thus allows us to estimate the
completeness-corrected LF in the face of incomplete information.  The
drawbacks to the forward approach are that there is no obvious way to
compute the uncertainties in the resulting LF and that some {\sl a
priori\/} information is needed about the true LF\@.  This forward
method will give different results depending on the initial guess at
$\phi(V)$ and the scheme used to adjust $\phi^{\prime}$ during the
iteration.  We found that, for our data, the LF is not dependent on
the initial guess at the form of the LF when $V \lesssim 25$
(approximately the level of the HB).  We adjusted $\phi^{\prime}(V)$
using $\phi^{\prime}_i(V) = \phi^{\prime}_{i-1}(V) + \Delta n(V)$ and
iterated $n^{\prime}(V) = {\bf P}\phi^{\prime}(V)$ until $\Delta n$
was minimized.

\begin{figure}
\plotone{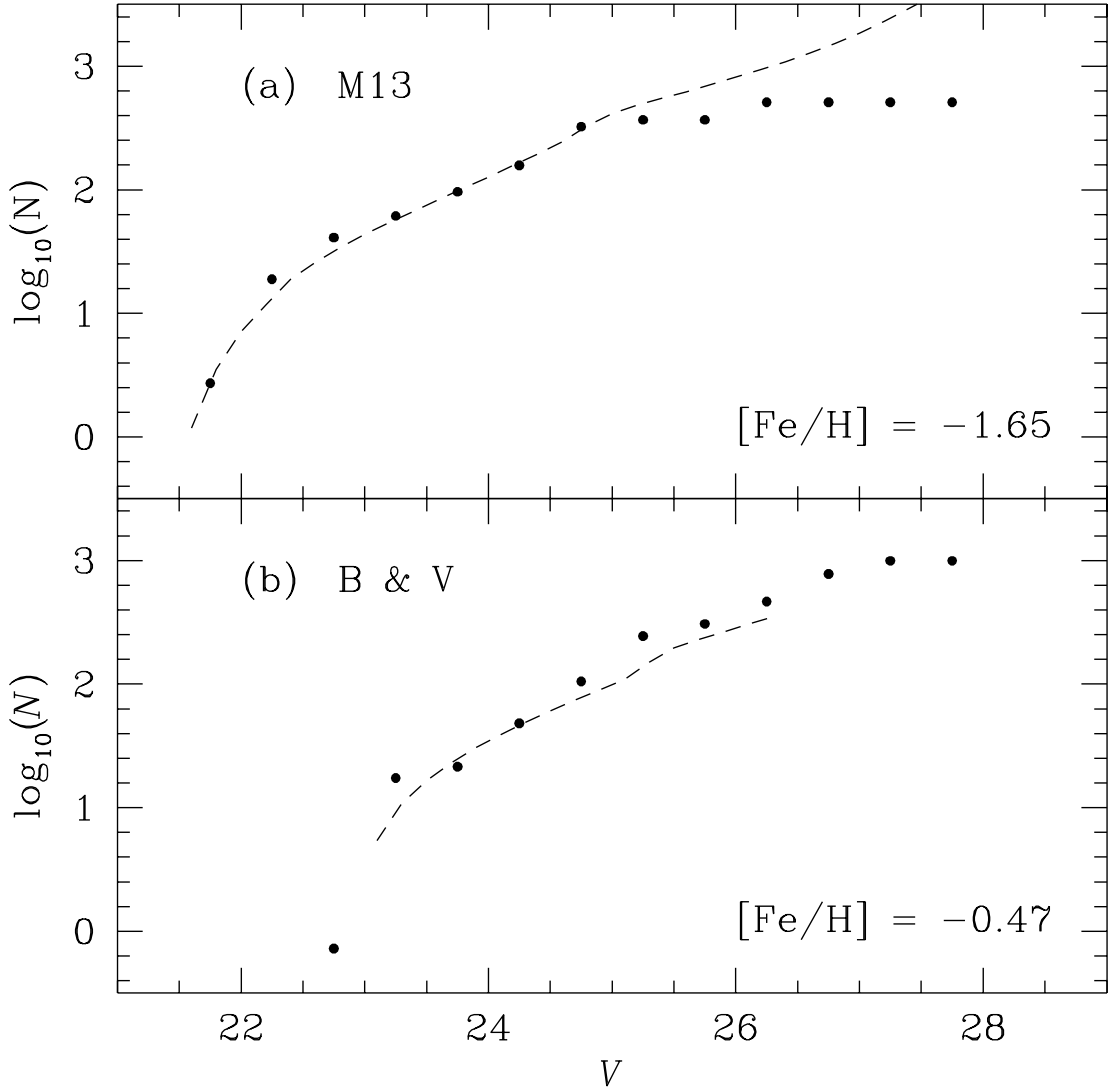}
\caption{The upper panel (a) compares the cumulative LF for
G302 (solid circles) to the cumulative LF of the metal-poor Galactic
GC M13.  Panel (b) compares the cumulative LF for G312 with a
theoretical LF for an old metal-rich GC\@.  The comparison LFs have
been scaled so that they contains the same number of stars at $V = 24$
as the M31 GC LFs do. \label{FIGURE:lfs_comp}}
\end{figure}

	Figure~\ref{FIGURE:lfs} shows the observed and
completeness-corrected, background-subtracted LFs for G302 and G312.
These LFs are reliable for $V \lesssim 25$.  In
Figure~\ref{FIGURE:lfs_comp}a we compare the cumulative LF for G302
with that of the metal-poor ($\FeH = -1.65$) Galactic GC M13 (Simoda
\& Kimura \markcite{SK68}1968).  In Figure~\ref{FIGURE:lfs_comp}b we compare
the cumulative GC for G312 to a theoretical LF for an old ($t_0 = 14$
Gyr) GC with $\FeH = -0.47$ and $\OFe = +0.23$ from Bergbusch \&
VandenBerg \markcite{BV92}(1992).  Stellar evolution models do not
show any relationship between the slope of the LF of the upper RGB and
either the iron abundance or the age of the stars (Bergbusch \&
VandenBerg \markcite{BV92}1992) so Figure~\ref{FIGURE:lfs_comp} can
not be used to determine the ages or the iron abundances of G302 or
G312.  There is a general agreement between the shape of the G312 LF
and the shape of the theoretical metal-rich LF, as well as between the
shape of the G302 LF and the metal-poor comparison LF, which suggests
that our star counts are reasonably complete down to the level of the
HB ($V \simeq 25$).


\section{Structure\label{SECTION:structure}}

\subsection{Michie--King Models\label{SECTION:MK}}

	We determined the tidal radii by fitting MK models to the
background-subtracted surface brightness profiles of each GC\@.  The
unresolved background light was measured by masking out a circle of
radius 300 pixels ($\sim 3 r_t$) centered on the GC and fitting a
plane to the remaining area on the WF3 CCD\@.  We found that
subtracting a constant sky level resulted in residual variations of
$\lesssim 1$\% across the WF3 CCD for both the G302 and the G312
fields.  We found no evidence for intensity or color gradients in the
unresolved background light.

	Surface brightness profiles for each GC were obtained using
the IRAF\footnote{Image Reduction and Analysis Facility (IRAF), a
software system distributed by the National Optical Astronomy
Observatories (NOAO).}  implementation of the {\sc stsdas} task {\sc
ellipse}.  This task uses Jedrzejewski's \markcite{J87}(1987)
isophote-fitting algorithm to fit elliptical isophotes to the observed
isophotes on the background-subtracted images.  This algorithm
determines the centers, ellipticities, orientations, and intensities
for a series of ellipses logarithmically spaced between $\sim$
0\farcs1 ($\sim$ 1 pixel) and $\sim$ 20\farcs0 ($\sim$ 200 pixels).
The spacing between successive ellipses varies from significantly less
than the size of the WF3 seeing disc (FWHM $\sim 0\farcs2$) near the
center of each GC to $\sim$10 times the size of the seeing disc beyond
the tidal radius of the GC\@.  The properties of the ellipses are not
independent of each other for radii of less than approximately eight
times the size of the seeing disc (\eg Schweizer \markcite{S79}1979).
Since the spacing between ellipses is logarithmic the degree of
correlation will vary with the distance from the center of the GC\@.
There is an additional correlation caused by the overlap in the
subsets of pixels being used to compute successive best-fitting
ellipses.  In light of these effects the computed uncertainties in the
surface brightnesses, ellipticities, and orientations of the
individual fitted ellipses should be treated as guides to the
reliability of the individual values relative to the values for the
other fitted ellipses for that GC and {\sl not\/} taken as statistically
rigorous estimates.  In the inner regions of each GC ($r \lesssim
1$\farcs0) the large gradient in surface brightness, and the small
number of pixels, make the fitted ellipses less reliable than those
fit further out.  Finally, we extracted the one-dimensional surface
brightness profile for each GC along the effective radius axis of each
GC\@.\footnote{The effective radius, $\reff$, of an ellipse with
semi-major and semi-minor axes $a$ and $b$ respectively is defined by
$\pi \reff^2 \equiv \pi a b$.  The effective radius axis is either of
the two axes along which the perimeter of the ellipse intersects the
perimeter of a circle with radius $\reff$.}.

	We used CERN's {\sc minuit} function minimization package to
fit single-mass one-dimensional seeing-convolved MK models to the
surface brightness profiles for each GC\@.  The fitting was done by
simultaneously adjusting all the free parameters until the
$\chi^2_\nu$ statistic between the model and the weighted surface
brightness profile was minimized.  After some experimentation we
adopted a weighting scheme of $w_i = (1/\sigma_i)^2$ where $\sigma_i$
is the RMS scatter in the isophotal intensity about the $i^{\rm th}$
fitted isophotal ellipse.  We tested this weighting scheme by fitting
MK models to a series of surface brightness profiles created by
drawing points from a MK distribution and found good agreement between
the input and the fitted models.  The exact form of the weighting did
not significantly affect the values of the best-fitting model
parameters provided that the weights assigned to each data point were
reasonably indicative of the reliability of each point.

\begin{figure}
\plotone{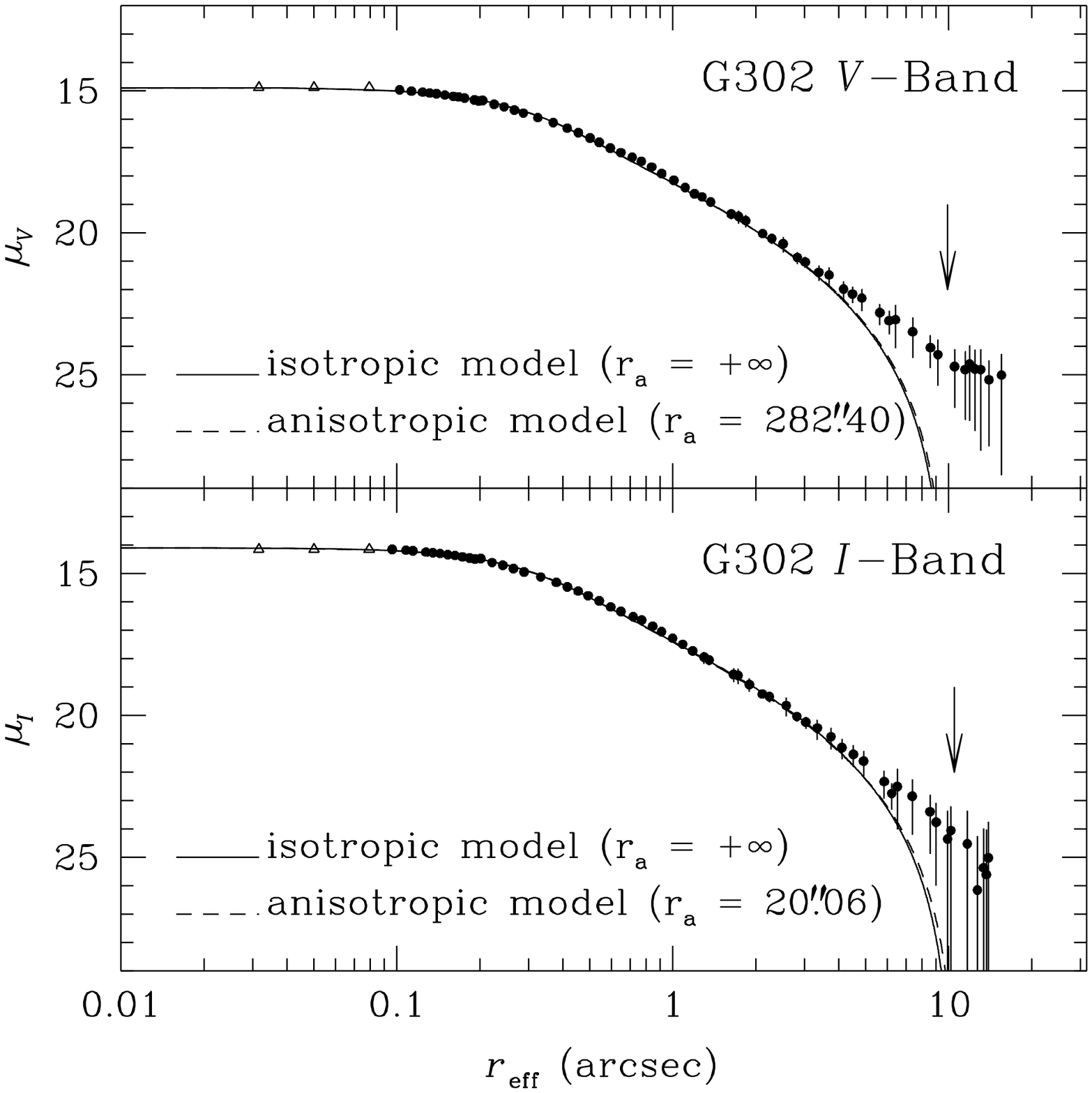}
\caption{The solid points represent the observed surface brightness
profile of G302, while the open triangles in the inner $0\farcs1$
indicate the surface brightness corresponding to saturation for the
WF3 CCD\@.  The solid line is the best-fitting isotropic MK model and
the dashed line is the best-fitting anisotropic model.  The arrows
indicate the tidal radii for each model. \label{FIGURE:G302_sdp}}
\end{figure}

	Each MK model was convolved with the PSF for the WF3 CCD at
the center of the GC\@.  We determined the central potential, $W_0$,
King core radius, $r_c$, tidal radius, $r_t$, concentration, $c
\equiv r_t/r_c$, anisotropy radius, $r_a$, and half-mass radius,
$r_h$, for G302 and G312 in the \V- and \I-bands.
Table~\ref{TABLE:1d_models} gives the parameters of the best-fitting
MK models for each GC\@.  The \V- and \I-band surface brightness
profiles, along with the best-fitting MK models, are shown in
Figures~\ref{FIGURE:G302_sdp}~and~\ref{FIGURE:G312_sdp}.  The profiles
for both GCs are well fit by isotropic MK models out to $\sim
5\arcsec$.  Beyond this distance from the center of the GC there is
excess light over what is predicted by isotropic ($r_a = +\infty$) MK
models.  In order to test if the excess light is due to velocity
anisotropy in the GCs we fit anisotropic MK models to each GC\@.  The
results are listed in Table~\ref{TABLE:1d_models}.  In all cases the
anisotropic models yield slightly larger $\chi^2_{\nu}$ values than
the corresponding isotropic models, suggesting that the isotropic
models are formally better fits to the data than are the anisotropic
models.  Furthermore, the best-fitting anisotropy radii are greater
than the fitted tidal radii for each GC, which is physically
meaningless.  Therefore, we conclude that the overabundance of light
at large radii is not due to velocity anisotropy.  This supports the
hypothesis that these GCs are surrounded by extended halos of tidally
stripped stars.

	{\sc Minuit} returns formal uncertainties of $\sim$ 5 to 10\%
on the MK parameters.  However, {\sc minuit} was generally unable to
compute a fully accurate covariance matrix for the fitted parameters
so {\sc minuit}'s uncertainty estimates are not reliable.  We
estimated the uncertainties in our fits by generating a series of
artificial GCs and fitting seeing-convolved MK models to them in
exactly the same way that we did to our data.  We found that {\sc
minuit} tended to underestimate the uncertainties in our fits by a
factor of between 1.5 and 2.  This suggests the true 1-$\sigma$
uncertainties in the parameters are between $\sim$10 and 15\% of the
best-fit values of the parameters.

\subsection{Color Gradients\label{SECTION:colors}}

	The mean color for $\reff \le 2\farcs5$ for G312 is
$\langle(V-I)_0\rangle = 1.07$ while for G302 $\langle(V-I)_0\rangle =
0.83$.  These colors are consistent with G302 having a metal-poor
stellar population and G312 having a metal-rich stellar population.

	Figure~\ref{FIGURE:colours} shows the dereddened $(V-I)_0$
color profiles of G302 and G312.  These profiles were obtained by
subtracting the isophotal surface brightnesses in $V$ and $I$ then
dereddening the resulting profiles.  The error bars represent the
1-$\sigma$ standard error in the color.  The inner portion of the G302
profile is truncated at $\reff \sim 0\farcs1$ ($=$ 1 pixel) since the
innermost pixels ($r \le 1$ pixel) of our images of G302 were slightly
saturated.  Individual stars can be resolved to within $\reff
\simeq 2\farcs5$ of the center of the GC and the isophotes become less
certain when the light becomes resolved into individual stars.
Artificial star tests suggest that the large scatter in the color
profiles in the outer regions of G302 and G312 represent stochastic
star placement within the GCs, not intrinsic color gradients.  In the
inner $2\farcs5$ the color gradients are $d (V-I)_0 / d \reff = -0.028
\pm 0.007$ for G302 and $d (V-I)_0 / d \reff = -0.090 \pm 0.007$ for
G312.  To determine if these color gradients are due to chance we
performed 10,000 bootstrap resamplings of the color data for each GC
and computed the color gradients for each resampled data set.  This
analysis gave a 2.1255\% probability that the observed color gradient
(for $\reff \le 2\farcs5$) in G302 is due to chance and a 0.0005\%
probability that the observed color gradient (for $\reff
\le 2\farcs5$) in G312 is due to chance.

\subsection{Ellipticities\label{SECTION:ell}}

	Figures~\ref{FIGURE:G302_shape}~and~\ref{FIGURE:G312_shape}
show the ellipticity and orientation profiles for G302 and G312
respectively.  The error bars in
Figures~\ref{FIGURE:G302_shape}~and~\ref{FIGURE:G312_shape} should be
treated as guides to the reliability of the individual measurements
and {\sl not\/} taken as statistically rigorous estimates of the
uncertainties in each point.  In order to estimate the overall
projected ellipticity of each GC we determined the weighted mean
ellipticities, $\overline{\epsilon}$ and position angles,
$\overline{\theta_0}$ (in degrees east of north on the sky) using the
values for the individual isophotes between $\reff = 1\arcsec$ and
$\reff = 10\arcsec$.  The results are presented in
Table~\ref{TABLE:shapes}.  The uncertainties are standard errors in
the mean and $N$ is the number of isophotes used to compute
$\overline{\epsilon}$ and $\overline{\theta_0}$.

	The projected major axis of G302 points approximately towards
the center of M31.  The ellipticity of each GC is the same in the
\I-band as it is in the \V-band but the projected major axis of G302
in the \V-band is located $16\arcdeg \pm 3\arcdeg$ west of the
projected major axis in the \I-band.  This shift in orientation
between the two colors is visible at all radii in G302 and is likely
due to a small number of giant stars dominating the light from the
GC\@.  Our data show that the projected major axis of G302 in the
\V-band is oriented $20\arcdeg$ farther west from the \V-band
orientation obtained by Lupton \markcite{L89}(1989) ($\theta_0 =
0\fdg0 \pm 0\fdg6$), but that Lupton's orientation is only $4\arcdeg$
east of our \I-band orientation.  This supports the idea that the
difference between the \I- and \V-band orientations that we observed
is due to the different numbers of resolved stars in the different
bands and data sets.  Our ellipticity ($\epsilon = 0.195 \pm 0.012$)
is somewhat higher than that found by Lupton \markcite{L89}(1989)
($\epsilon = 0.15 \pm 0.02)$.  Figure~\ref{FIGURE:G302_shape} shows an
apparent decrease in the ellipticity of G302 between $\reff \sim
1\arcsec$ and $\reff \sim 10\arcsec$.  The slope of $\epsilon(\reff) =
m \reff + b$ is $m = -0.0381 \pm 0.0383$.  In order to determine if
this slope represents a real decrease in ellipticity with radius we
performed 10,000 bootstrap resamplings of the ellipticity data between
$1\arcsec \le \reff \le 10\arcsec$ and computed the slope for each set
of bootstrapped ellipticities.  We found that there was a 22.34\%
chance of observing a slope at least as great as the slope seen in the
actual data simply by chance.  Therefore, we conclude that there is no
evidence for a change in ellipticity with radius between $\reff =
1\arcsec$ and $\reff = 10\arcsec$ in G302.  It is not possible to say
if the apparent increase in ellipticity between $\reff \simeq
0\farcs3$ and $\reff \simeq 1\farcs0$ is real or due to the presence
of a small number of unresolved bright stars located within a few core
radii of the center of G302.  Since the stellar FWHM is $\sim
0\farcs2$, the chance location of a single unresolved bright star
within $0\farcs3$ of the center of G302 along the projected minor axis
of the GC could reduce the ellipticity of the fitted isophotes.

	G312 has a projected ellipticity of $\overline{\epsilon} =
0.065$ but there is considerably more scatter in the fitted position
angles (the standard deviation is $\sigma \sim 27\arcdeg$) than in the
G302 position angles ($\sigma \sim 10\arcdeg$).  This suggests that
the observed projected ellipticity of G312 is primarily due to the
stochastic placement of stars within G312, variations in the stellar
background of the M31 halo, and the intrinsic precision of the {\sc
ellipse} algorithm.  Running the {\sc ellipse} task on a series of
circular artificial MK GC models which were convolved with the WF3 PSF
suggests that seeing alone can produce ellipticities of between 0.01
and 0.02 at radii of between approximately $0\farcs5$ and $2\arcsec$.
This agrees with the fluctuations in ellipticity seen in
Figure~\ref{FIGURE:G312_shape} at these radii.  While seeing has very
little effect on the observed ellipticity beyond $\reff \sim
2\arcsec$, artificial star tests indicate that stochastic star
placement can introduce an uncertainty in the ellipticity of between
$\pm 0.01$ and $\pm 0.05$.  In light of this we believe there is no
evidence for G312 being elliptical in the plane of the sky.  It is,
however, possible that G312 is elongated along the line of sight.


\section{Extended Stellar Halos\label{SECTION:tails}}

	GAF have observed an excess of resolved and unresolved stars
beyond the formal MK tidal radii of several GCs in M31, as would be
expected if the stars which evaporated from these GCs remained in
extended halos around those GCs.  Another test for extended halos of
unbound stars is to identify an asymmetrical overdensity of stars in
two dimensions beyond the formal tidal radius.  This has been done for
some Galactic globular clusters (see Grillmair \etal
\markcite{GF95}1995) using multiple Schmidt photographic plates for
each GC\@.  However, it has not been attempted for the M31 GCs, where
the GC, and the background, can be imaged on a single CCD\@.  This
eliminates possible systematic effects arising from comparing star
counts across multiple fields.

	To search for extended halos around G302 and G312 we looked at
the two-dimensional distribution of stars beyond the tidal radii of
each GC\@.  The total background stellar densities around each GC were
small ($\Sigma_{\rm bkgd} = 0.6191 \pm 0.0099 \; {\rm
stars}/\sq\arcsec$ for G302 and $\Sigma_{\rm bkgd} = 0.1688 \pm 0.0052
\; {\rm stars}/\sq\arcsec$ for G312).  We rebinned the stellar positions
into ``super-pixels'' so there were enough stars in each bin to ensure
that the Poisson fluctuations in a bin would be small compared to the
total number of stars in that bin.  In order to compute the optimum
binning size we assumed that all the stars on the WF3 CCD were part of
the GC and that the GC could be approximated by a Gaussian
distribution of stars.  It can be shown (Heald \markcite{H84}1984)
that for any Gaussian distribution the bin size, $\delta x$, which
maximizes the signal-to-noise in each bin, is given by $\delta x \sim
(20/N)^{1/5} \sigma$ where $\sigma$ is the standard deviation of the
Gaussian and $N$ is the size of the sample.  This technique provides a
balance between the need to keep the bin sizes small compared to the
width of the Gaussian, and the need to have a large number of
data-points in each bin to reduce Poisson noise in the bin.  Since we
are searching for overdensities of stars beyond the MK tidal radii of
the GCs the assumption that a GC can be approximated by a Gaussian
will result in an underestimate of the bin size needed to maximize the
signal-to-noise.  We found that bin sizes of 32 pixels ($\simeq$
3\farcs2), approximately two to three times the computed optimal size
for Gaussians, provided the best signal-to-noise for our data.  We
smoothed the resulting binned data by convolving it with a unit
Gaussian with a dispersion equal to the bin size ($\sigma =
3\farcs2$).  The resulting stellar number density distribution for
G302 is shown in Figure~\ref{FIGURE:G302_tail} and the stellar number
density distribution for G312 is shown in
Figure~\ref{FIGURE:G312_tail}.

	Figure~\ref{FIGURE:G302_tail} shows an asymmetric overdensity
of stars around G302 extending to at least twice the MK tidal radius.
The coherence of the isodensity contours which lie beyond the tidal
radius, but inside the background contour, suggests that the observed
overdensities are not an artifact of the binning, smoothing, or
contouring processes.  The surface brightness profile of G302 (see
Figure~\ref{FIGURE:G302_sdp}) light beyond the tidal radius ($r_t \sim
10\arcsec$) of the GC is in agreement with the asymmetric halo of
stars seen in Figure~\ref{FIGURE:G302_tail}.  The surface brightness
profile of G312, however, is consistent with there being no light
beyond the tidal radius ($r_t \sim 10\arcsec$).  This is reflected in
the lack of an extended halo of stars in
Figure~\ref{FIGURE:G312_tail}.  The fact that an overdensity of stars
is seen beyond the tidal radius of G302, but not beyond the tidal
radius of G312, suggests that the overdensity is real and not an
artifact of the analysis.

	In order to determine if the observed halo surrounding G302 is
real we applied our contouring procedure to the star counts from the
WF2 and WF4 images in the G302 and G312 fields.
Figure~\ref{FIGURE:bkgd_tail} shows that no structures of comparable
size to the halo around G302 are seen in any of the four background
fields.  To test whether a random distribution of stars could give
rise to coherent structures which could be mistaken for an extended
halo we constructed a series of random star fields with number
densities comparable to those found in the G302 and G312 background
fields.  None of these artificial star fields showed evidence for
coherent sub-structure.  We also used MK surface density profiles to
place artificial stars onto randomly generated star fields to see if
the presence of a GC would bias our binning/smoothing/contouring
process in favor of finding structure beyond the tidal radius when
none was really present.  None of these Monte-Carlo images showed any
evidence for extended halos.

	In order to quantify the orientation of the halo around G302
we computed the second moment of the distribution of stars beyond the
fitted tidal radius.  Only stars with $10\arcsec \le r \le 35\arcsec$
from the center of the GC were used to ensure that neither the GC or
the edges of the CCD biased our sample.  We computed the statistic
$\zeta_2 = \sqrt{{1 \over N}\sum_{i=1}^N y_i^2}$ where $N$ is the
number of stars beyond the tidal radius and $y_i$ is the distance from
the $i^{\rm th}$ star to an arbitrary axis of symmetry for the GC\@.
This statistic corresponds to the root-mean-square distance between
the stars and the axis of symmetry so the value of $\zeta_2$ will be
at its minimum when the assumed axis of symmetry coincides with the
true major axis of the distribution.  The $\zeta_2$ values for G302
and G312, assuming a range of angles of symmetry, are shown in
Figure~\ref{FIGURE:zeta}.  A second statistic, $\eta \equiv \zeta_{2,
\rm max} - \zeta_{2, \rm min}$, gives a measure of the degree of
symmetry in the distribution of stars about the center of the GC\@.
The value of $\eta$ for a distribution depends on the shape of the
distribution and the number of stars present.  For a given
distribution, the greater the value of $\eta$ measured the greater the
deviation from a circular distribution of stars.  For a perfectly
circular distribution, $\zeta_2$ would be the same for all angles
resulting in $\eta = 0$.  In order to determine the probability of
getting the observed value of $\eta$ by chance from a uniform circular
halo of stars around a GC we generated 10,000 sets of stellar
coordinates and measured $\eta$ for each set.  The stellar coordinates
were randomly drawn from a uniform distribution of stars with the same
size and number density as the field around the GC\@.  This data
formed the cumulative probability distributions shown in
Figure~\ref{FIGURE:eta_prob}.  Since the stellar number density in the
G302 field is different from the stellar number density in the G312
field we computed separate probability distributions for the G302
field and the G312 field.

	For G302 this analysis yields $\eta = 7.4826$, which occurs at
a position angle for the major axis of the extended halo of
$83\arcdeg~( = -97\arcdeg)$ east of north.  The cumulative probability
distribution for $\eta$ in the G302 fields (see
Figure~\ref{FIGURE:eta_prob}a) shows that there is only a 1.84\%
chance of obtaining a value of $\eta \ge 7.4826$ from a uniform
distribution of stars.  In contrast, the WF2 and WF4 fields near G302
yielded $\eta$ values of 3.5530 and 4.6618 corresponding to 38.47\%
and 20.62\% chances respectively of occurring by chance.  The
orientation of the extended halo around G302 is $10\arcdeg$ to the
west of the projected major axis of G302 as observed in the \V-band
and $26\arcdeg$ west of the projected major axis for the GC as
observed in the \I-band.  In light of our inability to find coherent
overdensities in any of the background images or simulated images, the
results of our moment analysis, and the observed overdensity of stars
beyond the tidal radius of the best-fitting MK model, we believe that
the extended asymmetric halo around G302 seen in
Figure~\ref{FIGURE:G302_tail} is a real feature of G302.

	Since the probability of observing a given value of $\eta$ by
chance depends on the number density of stars on the image we
performed a separate series of 10,000 Monte-Carlo simulations for the
G312 data (see Figure~\ref{FIGURE:eta_prob}b).  The resulting
probability distribution indicated that the observed value of $\eta
\ge 12.9960$ for the stars beyond the tidal radius G312 in the WF3
field had a 9.96\% chance of occurring in a randomly distributed set
of stars.  The WF2 and WF4 fields around G312 have $\eta$ values of
9.6342 and 9.1801 respectively, corresponding to 27.65\% and 30.73\%
probabilities of occurring at random.  This, and the lack of a
significant excess of stars beyond the tidal radius of the
best-fitting MK models, leads us to believe that G312 does not exhibit
any evidence for having an extended halo of stars.  We are not,
however, able to rule out the possibility that such a halo does exist
and is aligned along the line of sight.


\section{Mass Loss and the Orbit of G302\label{SECTION:mass_loss_and_orbit}}

\subsection{Mass Loss from G302\label{SECTION:mass_loss_G302}}

	Figures~\ref{FIGURE:G302_sdp}~and~\ref{FIGURE:G302_tail} show
that there is an excess of light and stars beyond the tidal radius of
the best-fitting MK model for G302.  By computing the amount of mass
beyond the tidal radius we can estimate the mass-loss rate for G302.
We converted our \V-band surface brightness profile for G302 to a
projected mass profile using a mean mass-to-light ratio of $\Upsilon_V
\equiv {\cal M} / L_V = 1.9$ Solar units (from Dubath \& Grillmair
\markcite{DG97}1997).  The total mass inside the MK tidal radius ($r_t
= 10\arcsec$) was ${\cal M} = (8.89 \pm 0.06) \times 10^5 \, {\cal
M}_{\sun}$.  We estimated the total mass of G302, by applying the
mass-to-light ratio to the total integrated \V-band magnitude of G302,
to be ${\cal M_{\rm tot}} = (9.52 \pm 0.24) \times 10^5 \, {\cal
M}_{\sun}$.  This gives a total mass of ${\cal M} = (0.63 \pm 0.25)
\times 10^5 \, {\cal M}_{\sun}$ beyond the formal tidal radius of
G302.  Mass-segregation within a GC can result in $\Upsilon_V$
increasing with radius so the total mass beyond the tidal radius that
we derive here will be a lower limit on the true mass in the extended
halo around G302.

	If we assume an age of $t_0 = 14$ Gyr and a constant rate of
mass loss for G302 then the mass-loss rate required for the observed
amount of mass to escape beyond the MK tidal radius is $\dot {\cal M}
= 4500 \pm 1800 \, {\cal M}_{\sun}$/Gyr.  The relaxation time at the
half-mass radius for a GC is given by (Spitzer \markcite{Sp87}1987)

\begin{equation}
t_{r,h} = 8.933 \times 10^5 ({\rm yr})
             {{\cal M}_{\rm cl}^{1/2} \, r_h^{3/2}  \over
               \langle{\cal M}_{\star}\rangle \ln(0.4 N_{\star})},
\end{equation}

\noindent
where ${\cal M}_{\rm cl}$ is the total mass of the GC in Solar masses,
$r_h$ is the half-mass radius of the GC in pc, $\langle{\cal
M}_{\star}\rangle$ is the mean stellar mass in the GC in Solar masses,
and $N_{\star} \equiv {\cal M}_{\rm cl} / \langle{\cal
M}_{\star}\rangle$ is the estimated number of stars in the GC\@.
Following Djorgovski \markcite{D93}(1993) we adopted $\langle{\cal
M}_{\star}\rangle = 1/3 {\cal M}_{\sun}$.  The half-mass radius
obtained by fitting isotropic MK models to G302 is $r_h = 1.9 \pm 0.1$
pc which gives a half-mass relaxation time of $t_{r,h} = 0.49 \pm
0.04$ Gyr.  This yields a projected escape rate of $\dot r = (2.3
\pm 0.9) \times 10^{-3}$ per half-mass relaxation time.

	Converting observed surface brightnesses to a deprojected mass
profile involves solving an Abel integral which contains the radial
derivative of the surface brightness distribution.  Since the surface
brightness data for G302 contains noise this inversion is inherently
unstable.  Therefore, we chose to project the theoretical escape rates
of Oh \& Lin \markcite{OL92}(1992) into the plane of the sky before
comparing them to our observed escape rate.  The Oh \& Lin
\markcite{OL92}(1992) evaporation rates for isotropic GCs with ages of
$t_0 \sim 30 t_{r,h}$ in the Galactic potential then become $\sim
10^{-3}$ to $10^{-2}$ per relaxation time.  This is comparable to the
escape rate inferred from our data.  This result further strengthens
our claim that we have detected an extended halo of unbound stars
around G302.

\subsection{Mass Loss from G312\label{SECTION:mass_loss_G312}}

\begin{figure}
\plotone{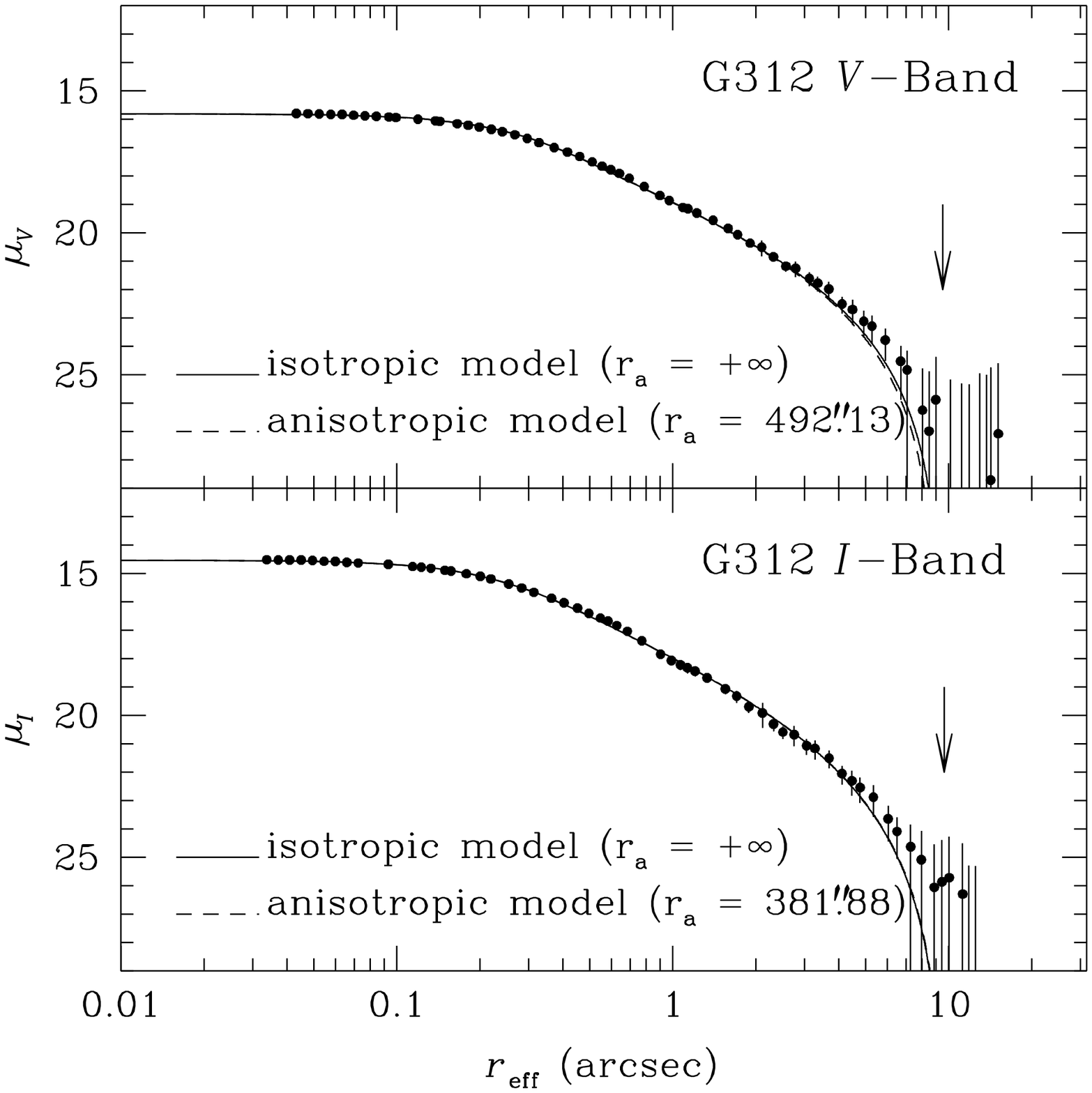}
\caption{The solid points represent the observed surface brightness
profile of G312.  The solid line is the best-fitting isotropic MK
model and the dashed line is the best-fitting anisotropic model.  The
arrows indicate the tidal radii for each model.
\label{FIGURE:G312_sdp}}
\end{figure}

\begin{figure}
\plotone{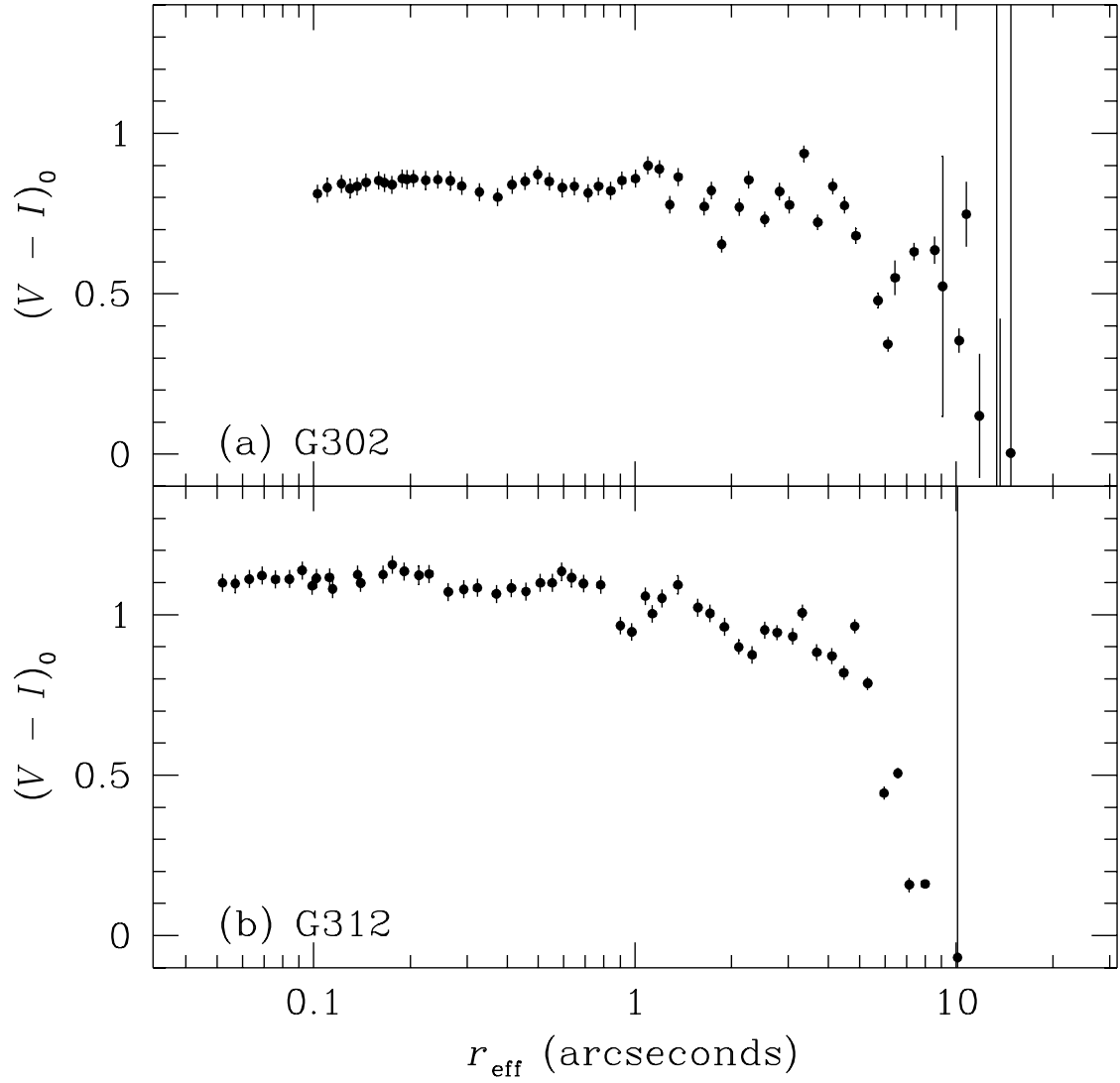}
\caption{This figure shows the integrated color profiles for G302 and
G312. \label{FIGURE:colours}}
\end{figure}

\begin{figure}
\plotone{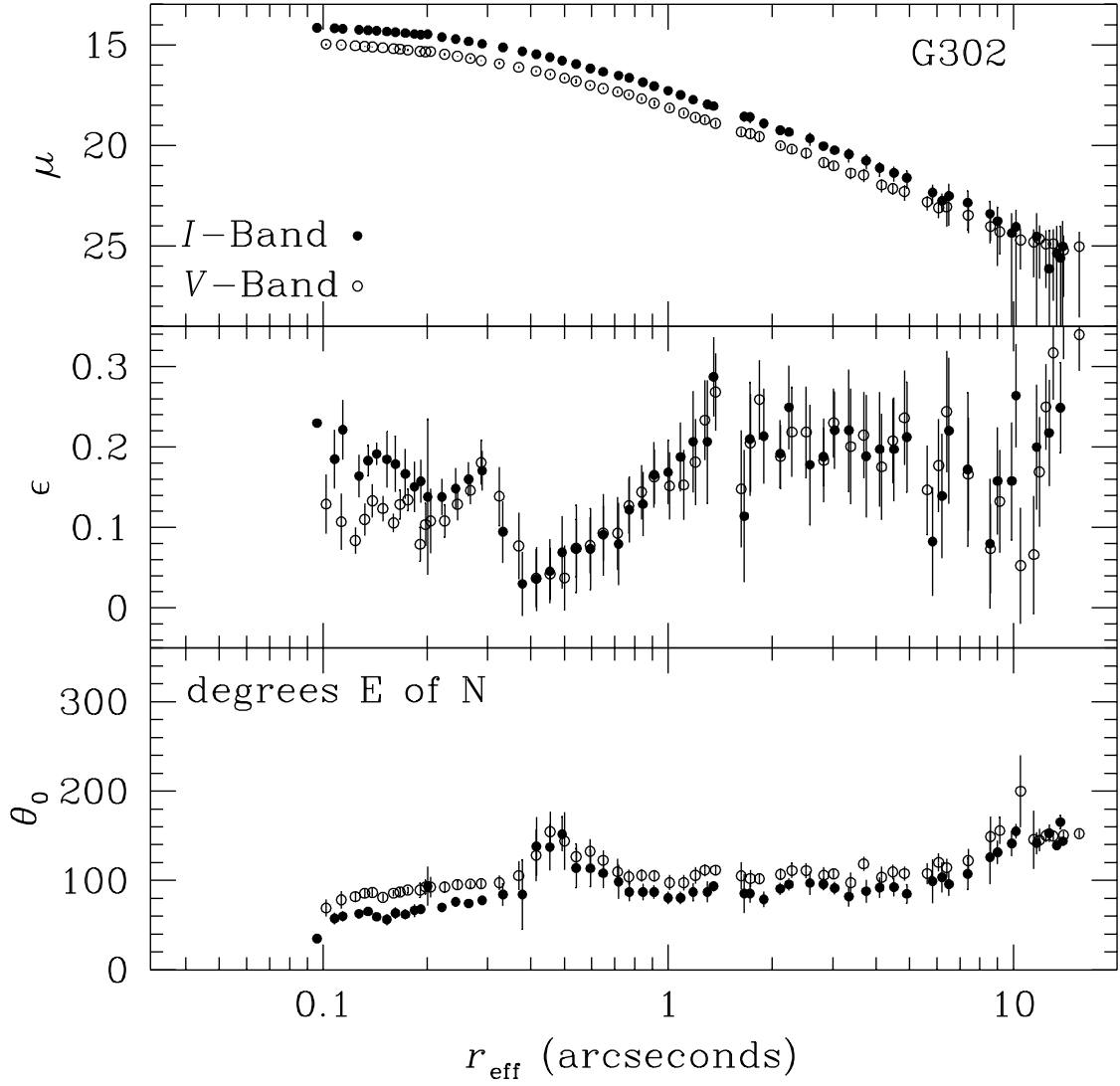}
\caption{This figure shows the calibrated surface brightness profiles,
$\mu$, ellipticity profiles, $\epsilon$, and orientation profiles,
$\theta_0$, for G302 in the \V- and \I-bands. \label{FIGURE:G302_shape}}
\end{figure}

\begin{figure}
\plotone{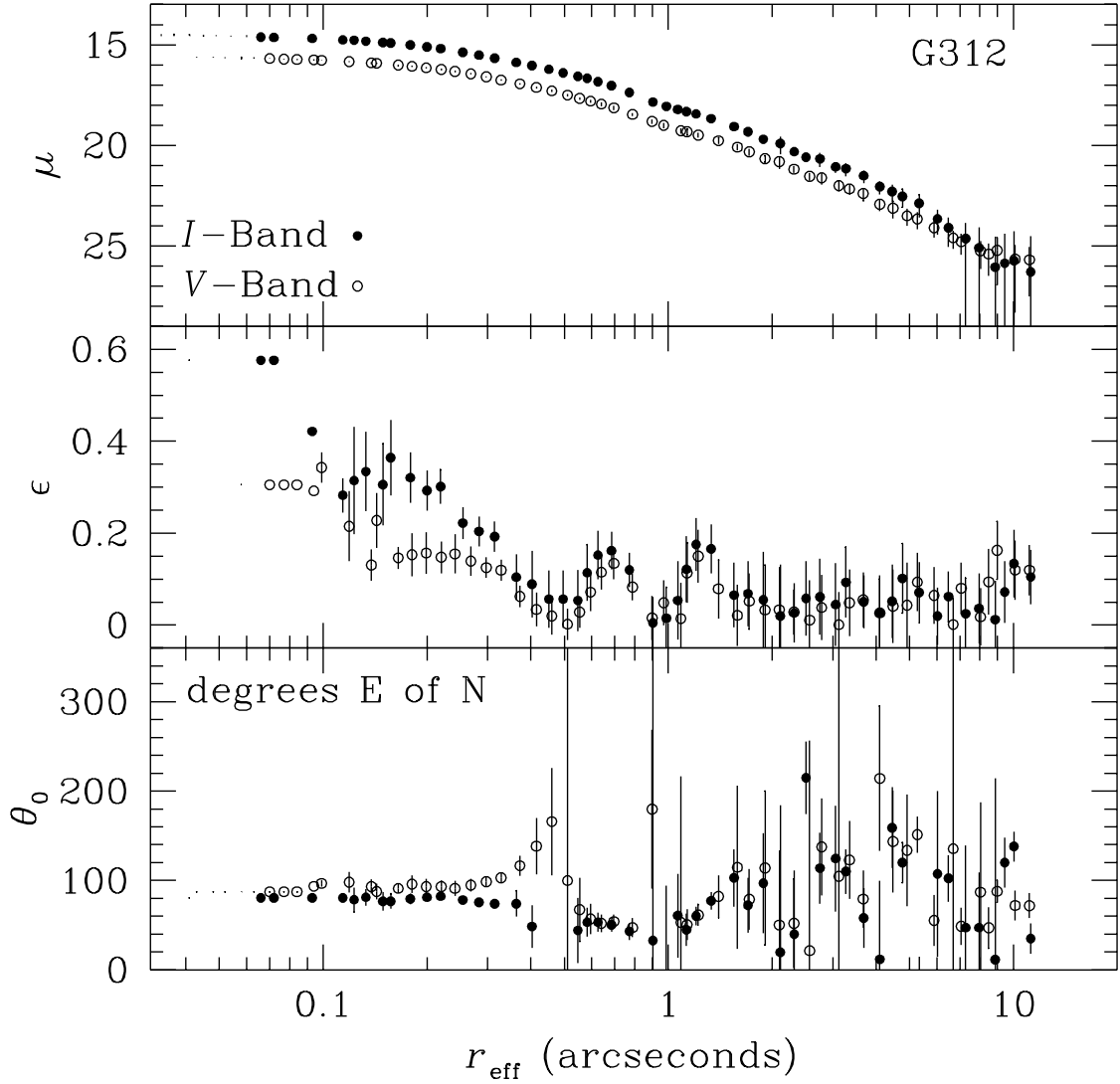}
\caption{This figure shows the calibrated surface brightness profiles,
ellipticity profiles, and orientation profiles for G312 in the \V- and
\I-bands. \label{FIGURE:G312_shape}}
\end{figure}

\begin{figure}
\plotone{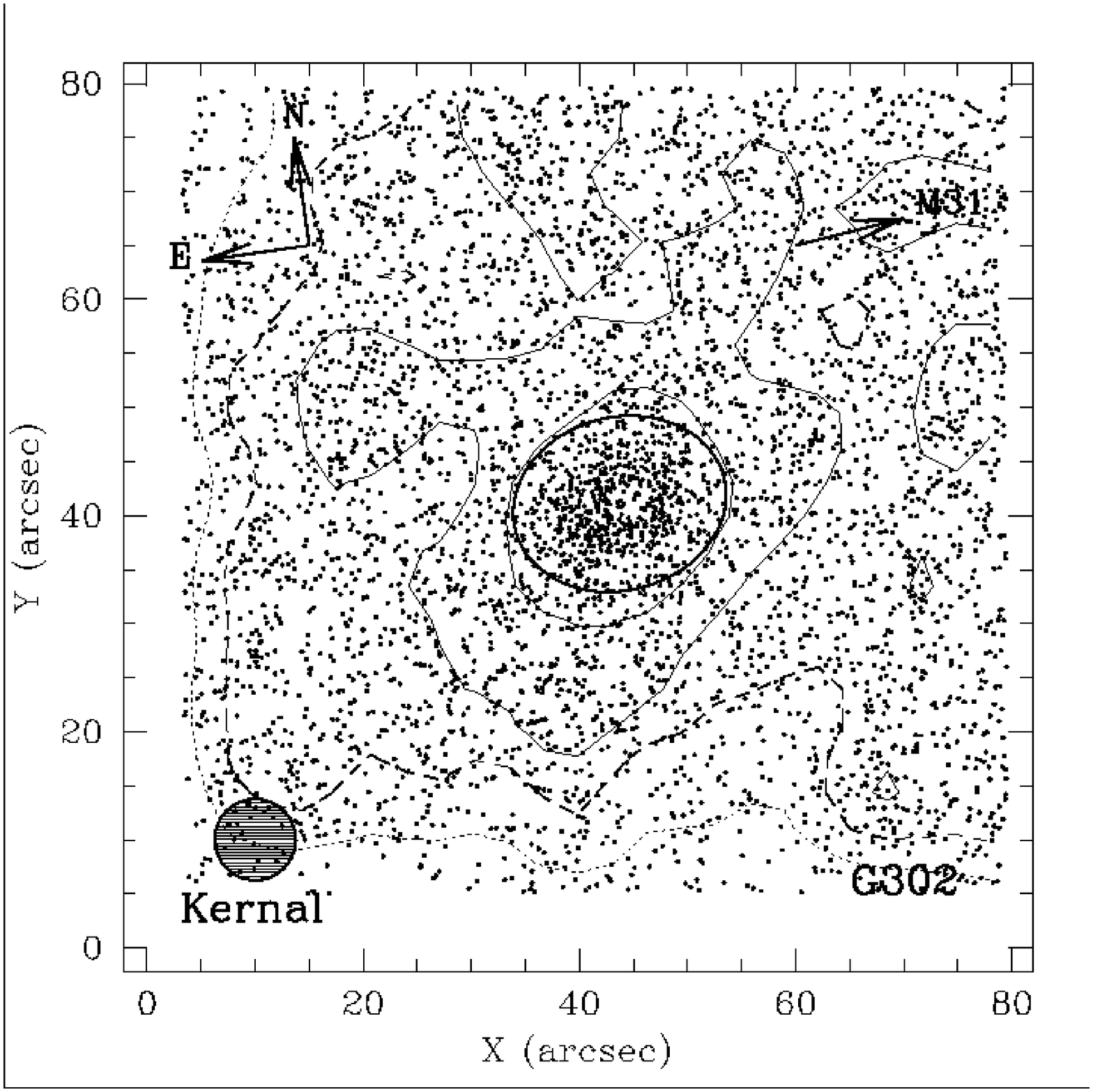}
\caption{The solid ellipse has an effective radius equal to the
fitted MK tidal radius for G302 and is oriented in the same direction
as the GC\@.  The dashed contour represents the surface density of
stars in the M31 halo at the location of G302 ($= \Sigma_{\rm bkgd} =
0.6196$ stars/\sq\arcsec).  The solid contours are $\Sigma = 0.8, 1.2
{\rm stars}/\sq\arcsec$ and the dotted contour is $\Sigma = 0.4 {\rm
stars}/\sq\arcsec$.  The outermost contours are strongly influenced by
the size of the WF3 CCD\@.  The contours inside the fitted tidal radii
have not been plotted for clarity.  The arrow points towards the
center of M31.
\label{FIGURE:G302_tail}}
\end{figure}

\begin{figure}
\plotone{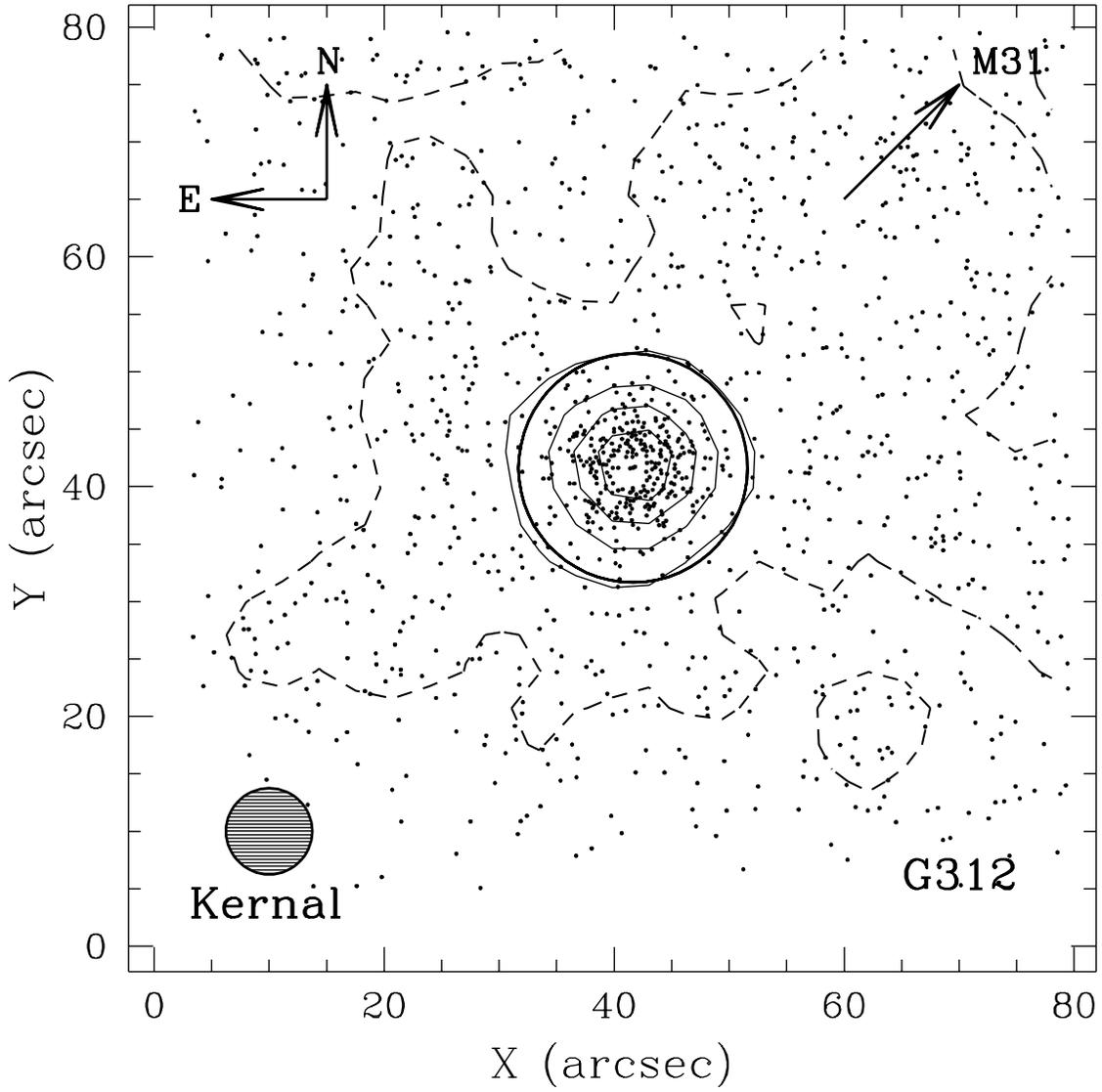}
\caption{The solid circle represents the fitted MK tidal radii of G312.
The dashed contour represents the surface density of stars in the M31
halo at the location of G312 ($= \Sigma_{\rm bkgd} = 0.1688$
stars/\sq\arcmin).  The solid contours are $\Sigma = 0.4, 0.8, 1.2,
1.6$ stars/\sq\arcmin.  The innermost contours have not been plotted
for clarity. \label{FIGURE:G312_tail}}
\end{figure}

\begin{figure}
\plotone{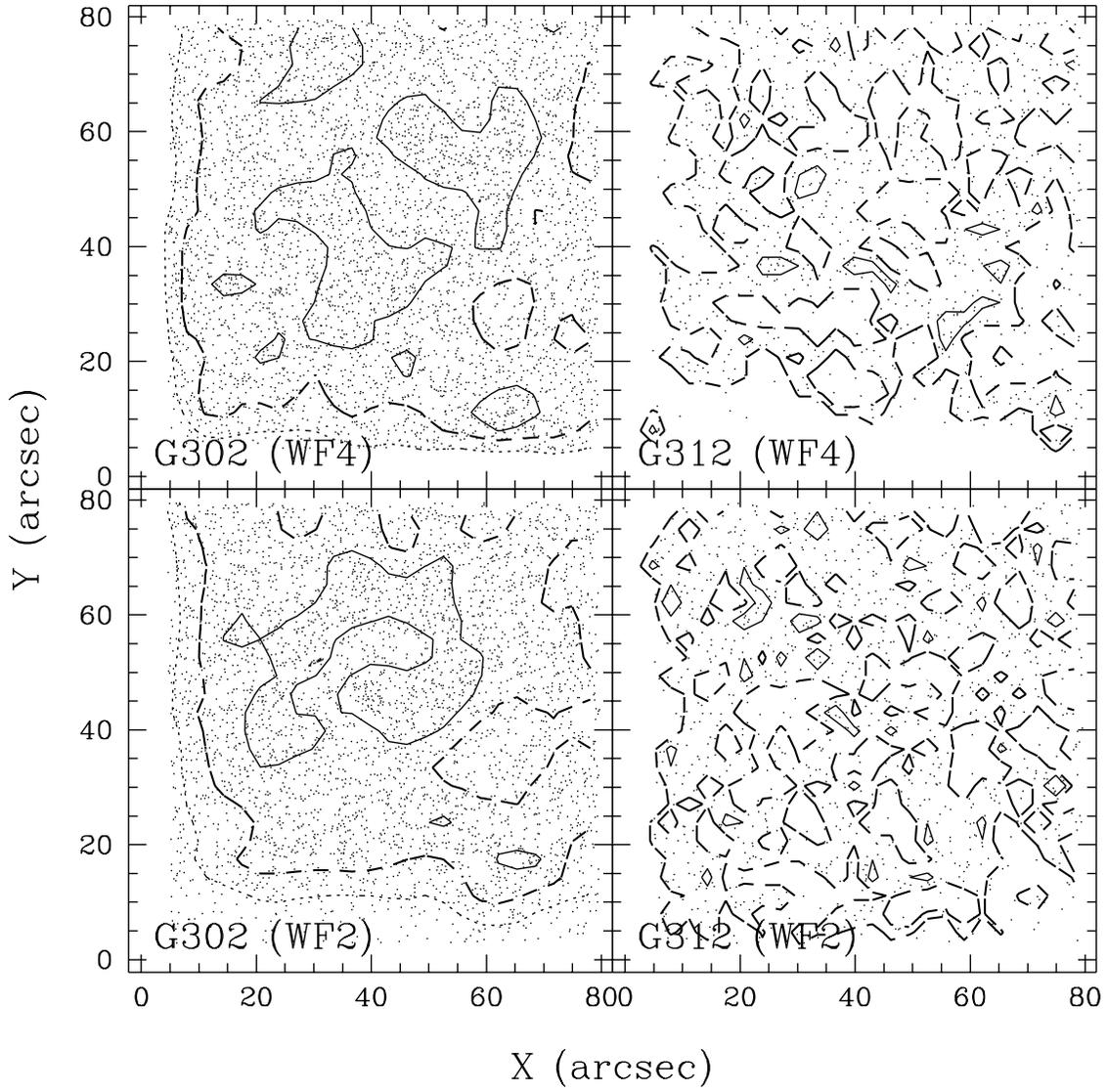}
\caption{This figure shows isodensity contours for the WF2 and WF4
fields for G302 and G312.  The contour intervals are the same as for
Figures~\ref{FIGURE:G302_tail}~and~\ref{FIGURE:G312_tail}.  The fields
in this figure are not oriented in the same direction as the WF3 CCDs
in Figures~\ref{FIGURE:G302_tail}~and~\ref{FIGURE:G312_tail}.
\label{FIGURE:bkgd_tail}}
\end{figure}

\begin{figure}
\plotone{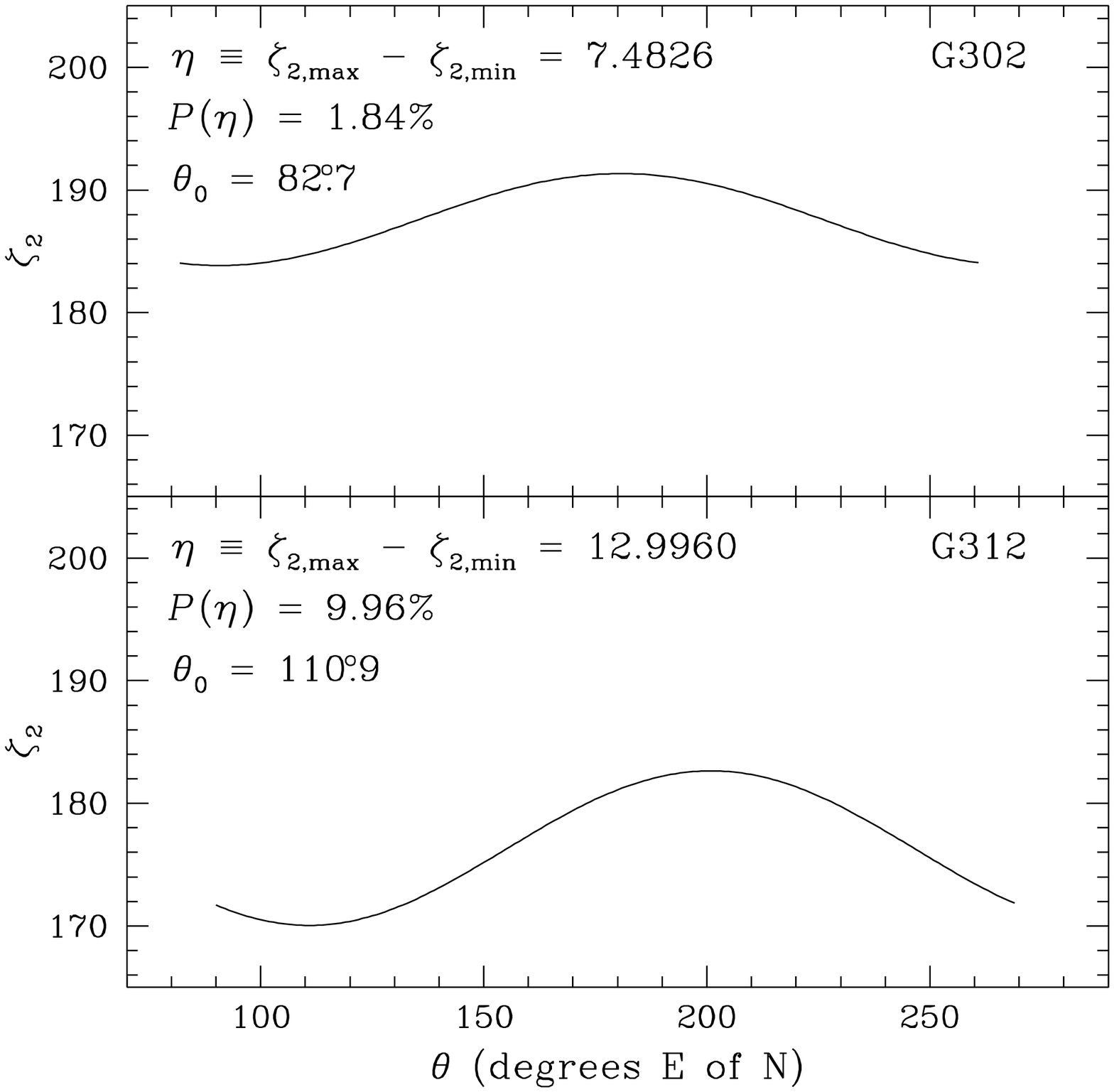}
\caption{The upper panel shows the value of $\zeta_2$ as a function
of position angle for G302, while the lower panel shows $\zeta_2$ as a
function of position angle for G312.  $\theta_0$ is the angle where
$\zeta_2$ is at its minimum, which corresponds to the position angle
(in degrees east of north on the sky) of the major axis of the GC\@.
$P(\eta)$ is the probability that the observed value of $\eta$ will
occur by chance in a uniform distribution of stars about the GC (see
Figure~\ref{FIGURE:eta_prob}). \label{FIGURE:zeta}}
\end{figure}

\begin{figure}
\plotone{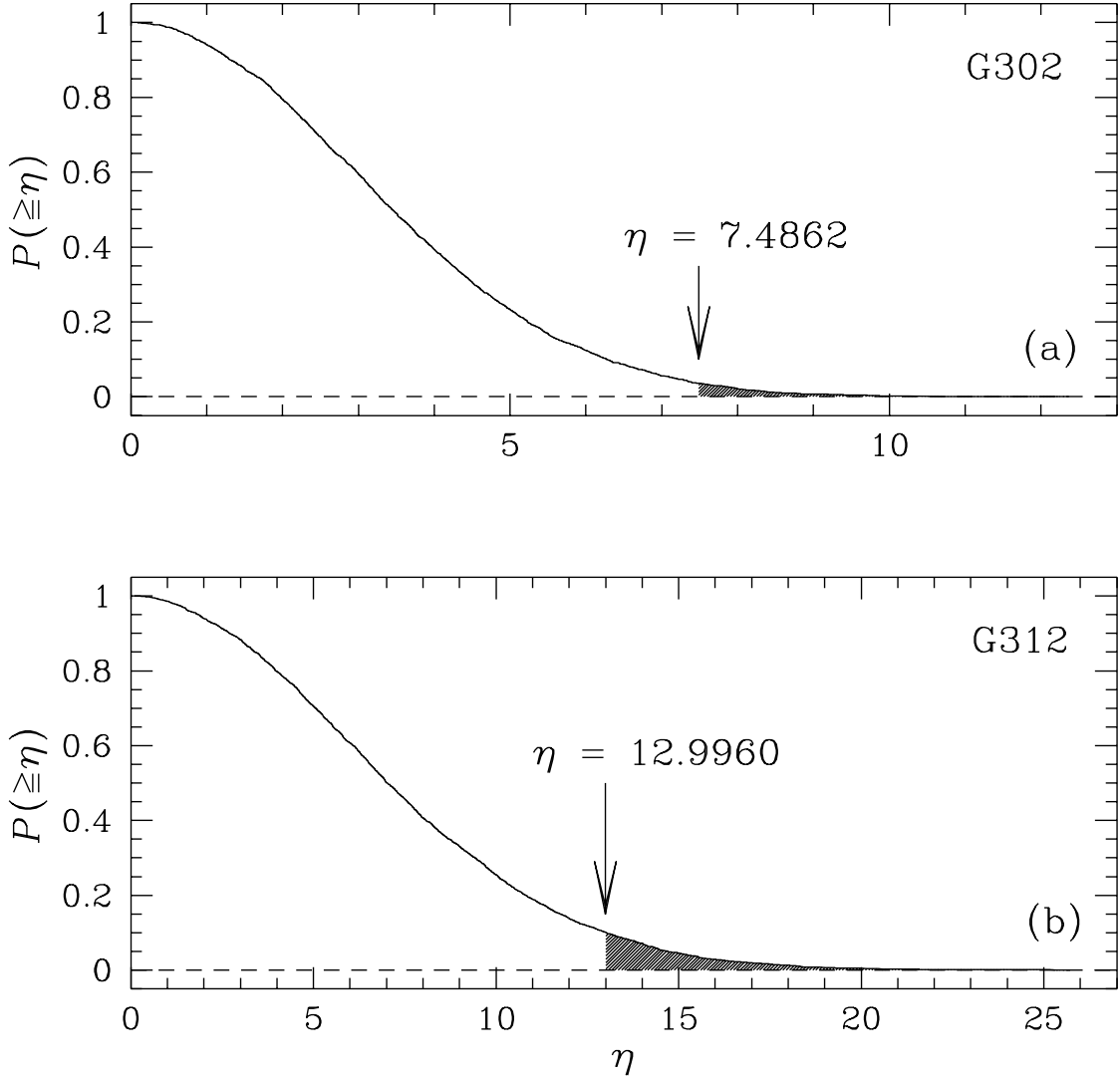}
\caption{This figure shows the cumulative probability distributions,
$P(\ge \eta)$, for observing a value $\ge \eta$ by chance if the
observed distribution of stars was drawn from a circularly symmetric
uniform distribution of stars.  Panel (a) shows $P(\ge \eta)$ for the
G302 field while panel (b) shows $P(\ge \eta)$ for the G312 field.
The two cumulative probability distributions are different because the
G302 and G312 fields contain different numbers of stars.  The shaded
areas under each curve represent the probabilities that values of
$\eta$ at least as large as those measured for G302 and G312 would be
observed by chance in a uniform distribution of stars about each GC\@.
\label{FIGURE:eta_prob}}
\end{figure}

	Using the methods described above we find that G312 has a mass
of ${\cal M} = (0.60 \pm 4.70) \times 10^4 {\cal M}_{\sun}$ beyond its
fitted isotropic MK tidal radius ($r_t = 9\farcs55$).  The large
uncertainty in the amount of mass beyond the tidal radius is due to
the large uncertainties in the surface brightness profile at large
radii.  Figure~\ref{FIGURE:G312_tail} does not show any evidence for
an extended halo of stars around G312 so it is likely that the light
seen beyond the fitted tidal radius in Figure~\ref{FIGURE:G312_sdp} is
due to low-level variations in the unresolved background light.
Alternately, it is possible that the use of multi-mass MK models would
result in a slightly larger tidal radius.  If we assume that the
observed excess of light in Figure~\ref{FIGURE:G312_sdp} is due to an
extended halo of stars, then the mass-loss rate required is $\dot
{\cal M} = 430 \pm 3360 \, {\cal M}_{\sun}$/Gyr, approximately 10\% of
the mass-loss rate from G302.  G312 has a half-mass relaxation time of
$t_{r,h} = 0.29 \pm 0.02$ Gyr, so the escape rate per relaxation time
is $\dot r = (0.38 \pm 2.95) \times 10^{-3}$.  This is consistent with
Oh \& Lin's \markcite{OL92}(1992) theoretical values.  However, the
large uncertainties in our photometry near the tidal radius of G312
(and thus the large uncertainty in our observed escape rate) suggest
that the observed escape rate is consistent with there being no
observable mass-loss from G312.

\subsection{The Orbit of G302\label{SECTION:orbit_G302}}

	HBK measured a heliocentric radial velocity of $-8 \pm 32$
km$\cdot$s$^{-1}$ for G302.  Correcting for Rubin \& Ford's
\markcite{RF70}(1970) 21 cm velocity for M31 ($v_{\rm M31} = -297$
km$\cdot$s$^{-1}$) gives G302 a radial velocity of $v = +289 \pm 32$
km$\cdot$s$^{-1}$ relative to the center of M31.  This is one of the
largest radial velocities observed for a GC in M31, suggesting that
almost all of G302's space velocity is along the line-of-sight.  This
can only occur when G302 is near its perigalactic passage.  Since very
little of G302's space velocity is in the plane of the sky, and G302
is near its perigalacticon, G302 must be at approximately the same
distance from us as is the center of M31.  This means that the
distance measured on the sky between G302 and the center of M31 is the
true separation; so the perigalactic distance for G302 is $\theta_p
\simeq 32\farcm1$ corresponding to $d_p \simeq 6.77$ kpc assuming a
distance modulus of $\mu_0 = 24.3$ for M31.

	In order to constrain the orbit of G302 we used a
three-component model for the M31 potential.  The disk was modeled by
a Miyamoto \& Nagai \markcite{MN75}(1975) potential, the bulge by a
Hernquist \markcite{H90}(1990) potential, and the halo by a spherical
logarithmic potential:

\begin{equation}
\Phi_{\rm disk} = {-G {\cal M}_{\rm disk} \over
                   \sqrt{R^2 + (a + \sqrt{z^2 + b^2})^2}},
\end{equation}

\begin{equation}
\Phi_{\rm bulge} = {-G {\cal M}_{\rm bulge} \over
                    r + c},
\end{equation}

\begin{equation}
\Phi_{\rm halo} = V^2_{\rm halo} \ln(r^2 + d^2).
\end{equation}

\noindent
We adopted ${\cal M}_{\rm disk} = 2.0 \times 10^{11} {\cal M}_{\sun}$,
${\cal M}_{\rm bulge} = 6.8 \times 10^{10} {\cal M}_{\sun}$, $a = 7.8$
kpc, $b = 0.31$ kpc, $c = 0.84$ kpc, $d = 14.4$ kpc, and $V_{\rm halo}
= 128$ km$\cdot$s$^{-1}$.  These parameters were chosen by scaling the
Galactic values to the observed mass and diameter of M31.

	Using this model, and the observed radial velocity and
perigalactic distance of G302, we found that the orbit of G302 has an
eccentricity of $e \equiv (d_a - d_p) / (d_a + d_p) = 0.65$ where
$d_a$ is the apogalactic distance (in kpc) for G302.  This value
assumes that all of the space velocity of G302 is along the
line-of-sight, and gives G302 an apogalactic distance of $\theta_a =
2\fdg5$ or $d_a = 31.5$ kpc.  If we allow 40\% of G302's space
velocity to be in the plane of the sky then the orbital parameters
become $e = 0.71$ and $d_a = 37.0$.  Since G302's radial velocity is
faster than 95\% of the radial velocities of other M31 GCs it is
unlikely that more than $\sim 5$\% of G302's space velocity is
tangential to the line-of-sight.  If this is the case, the orbital
eccentricity for G302 is $0.65 \lesssim e \lesssim 0.66$ and the
apogalactic distance is $d_a = 31.5$ kpc.  The period of G302's orbit
was defined to be twice the time required for the GC to move from its
perigalacticon to its apogalacticon.  If all of the GC's space
velocity is along the line-of-sight then the orbital period is $P =
0.42$ Gyr.  If 40\% of the space velocity is tangential to the
line-of-sight then $P = 0.52$ Gyr.


\section{Implications of the {\sl Hipparcos\/} Distance Scale\label{SECTION:Hipparcos}}

	Since this work was completed new parallax measurements from
the {\sl Hipparcos\/} satellite have been published which have led to
a revised estimate of the distance to M31.  The new distance modulus
is $\mu_0 = 24.77 \pm 0.11$ (Feast \& Catchpole \markcite{FC97}1997),
which corresponds to a distance of $d = 900 \pm 45$ kpc, a 25\%
increase over the distance that we adopted for M31.  If this distance
modulus is used then our results change in the following ways:

\begin{enumerate}

\item
Adjusting the fiducial RGB sequences in \sect~\ref{SECTION:cmds_G302}
and \sect~\ref{SECTION:cmds_G312} yields iron abundances of $\FeH =
-2.2$ for G302 and $\FeH = -0.7$ for G312.  The new G312 iron
abundance is consistent with the HBK spectroscopic iron abundance but
the new G302 value is $\sim 2.5$-$\sigma$ lower than the HBK or the de
Freitas Pacheco \markcite{dFP97}(1997) value.

\item
Chaboyer \etal \markcite{CD97}(1997) have derived a new relation,
$M_V({\rm RR}) = (0.23 \pm 0.04)(\FeH + 1.9) + (0.41 \pm 0.08)$,
between $\FeH$ and $M_V({\rm RR})$ which uses the {\sl Hipparcos\/}
parallaxes.  This relation, and the iron abundances and HB magnitudes
derived in \sect~\ref{SECTION:cmds_G302} and
\sect~\ref{SECTION:cmds_G312}, gives distance moduli of $\mu_0 = 24.51
\pm 0.12$ for G302 and $\mu_0 = 24.45 \pm 0.13$ for G312.  These
distance moduli would place G302 $\sim 100$ kpc in front of M31 if M31
is 900 kpc away.  This is inconsistent with the HBK radial velocity of
G302 and the orbital parameters determined in
\sect~\ref{SECTION:orbit_G302}.  The distance modulus for G312, and
the Feast \& Catchpole \markcite{FC97}(1997) distance to M31, puts
this GC $\sim 125$ kpc in front of M31.

\item
Adopting a distance modulus of $\mu_0 = 24.77 \pm 0.11$ would result
in the metal-rich GCs in Figure~\ref{FIGURE:vi-feh} being in better
agreement with the DCA relation and our extension to it.  However, the
metal-poor GCs would have bluer $(V-I)_{0,-3}$ colors which would
result in them falling to the left of the DCA relation.  This would
mean that the spectroscopic iron abundances of the metal-poor M31 GCs
have been systematically overestimated.  Since iron abundances are
more difficult to determine in spectra of metal-rich GCs it is more
likely that the $\FeH$ values for the metal-rich GCs have been
underestimated than the $\FeH$ values for the metal-poor GCs have been
underestimated.

\end{enumerate}

	In light of these results we find no reason to prefer the
$\mu_0 = 24.77 \pm 0.11$ distance modulus over the $\mu_0 = 24.3 \pm
0.1$ value.


\section{Conclusions\label{SECTION:conc}}

	We have used {\sl HST\/} WFPC2 photometry to construct deep
($V \simeq 27$) CMDs for two GCs in the halo of M31.  Both GCs appear
to have a single old population of stars similar to what is found in
Galactic GCs.  The shape of the RGB for G302 gives an iron abundance
of $\FeH = -1.85 \pm 0.12$, in agreement with the published values
obtained using spectroscopy.  For G312 we obtain $\FeH = -0.56 \pm
0.03$, which is somewhat more metal-rich than the spectroscopic value.
Neither GC shows any indication that there is a second parameter
acting upon their HB morphologies.

	Both GCs have MK tidal radii of $r_t \simeq 10\arcsec$, core
radii of $r_c \simeq 0\farcs2$, central concentrations of $c \simeq
1.7$, and half-mass radii of $r_h \simeq 0\farcs5$.  There is no
evidence for velocity anisotropy in either G302 or G312.  G302 has a
color of $(V-I)_0 = 0.83$ while the color of G312 is $(V-I)_0 = 1.07$.

	G302 has a projected ellipticity of $\epsilon = 0.195$ with
the major axis oriented approximately towards the center of M31.  This
GC has an excess of light beyond its formal tidal radius which is not
consistent with either an isotropic or an anisotropic MK model.  The
two-dimensional distribution of stars around G302 is consistent with
the presence of an extended halo extending to two to three times the
formal tidal radius from the cluster.  G312, on the other hand, has an
ellipticity of $\epsilon \simeq 0$.  Neither the integrated light, nor
the star counts, show any evidence for an extended halo.  It is
possible that such a tail does exist for G312, but is oriented along
the line of sight.

	We have estimated the projected mass-loss rate from G302 to be
$\dot {\cal M} = 4500 \pm 1800 \, {\cal M}_{\sun}$ per Gyr which
corresponds to a projected escape rate of $\dot r = (2.3 \pm 0.9)
\times 10^{-3}$ per half-mass relaxation time.  The projected
escape rate from G312 is $\dot r = (0.38 \pm 2.95) \times 10^{-3}$ per
half-mass relaxation time.  These are consistent with the escape rates
predicted by Oh \& Lin \markcite{OL92}(1992) although the large
photometric uncertainties near the tidal radius of G312 makes the
escape rate for this GC much less reliable than that for G302.


\acknowledgments

	This research is based on observations made with the NASA/ESA
{\sl Hubble Space Telescope\/} obtained at the Space Telescope Science
Institute.  STScI is operated by the Association of Universities for
Research in Astronomy Inc.\ under NASA contract NAS 5-26555.  Support
for this research was provided by operating grants to GGF and HBR from
the Natural Science and Engineering Research Council of Canada.  SH
would like to thank Peter Stetson for kindly making a copy of the {\sc
allframe} software, as well as \V- and \I-band PSFs for the WFC,
available.  We would also like to thank the anonymous referee who made
several useful comments.



\newpage

%
%

\begin{deluxetable}{rlcrcc}
\tablewidth{0 pt}
\tablecaption{Log of the Observations.\label{TABLE:obs_log}}
\tablehead
{\colhead{Field} &
 \colhead{Date} &
 \colhead{Filter} &
 \colhead{Exposure} \nl
 \colhead{} &
 \colhead{(1995)} &
 \colhead{} &
 \colhead{(s)}
}
\startdata
G302 & Nov.\ 5  & F555W  & 8 $\times$ 500  \nl
     &          &        & 2 $\times$ 160  \nl
     &          & F814W  & 7 $\times$ 500  \nl
     &          &        & 1 $\times$ 400  \nl
     &          &        & 1 $\times$ 160  \nl
G312 & Oct.\ 31 & F555W  & 8 $\times$ 500  \nl
     &          &        & 2 $\times$ 160  \nl
     &          & F814W  & 7 $\times$ 500  \nl
     &          &        & 1 $\times$ 400  \nl
     &          &        & 1 $\times$ 160  \nl
\enddata
\end{deluxetable}

%
%

\begin{deluxetable}{rrrccr}
\tablewidth{0 pt}
\tablecaption{Aperture Corrections.\label{TABLE:apcor}}
\tablehead
{\colhead{Field} &
 \colhead{CCD} &
 \colhead{Filter} &
 \colhead{$<{\rm ap} - {\rm PSF}>$} &
 \colhead{Slope} &
 \colhead{N}
}
\startdata
G302 &  PC & F555W & $-0.0226 \pm 0.0355$ &   0.992 &      12 \nl
     &     & F814W & $-0.0222 \pm 0.0218$ &   1.009 &      29 \nl
     & WF2 & F555W & $+0.0184 \pm 0.0216$ &   1.065 &      25 \nl
     &     & F814W & $+0.0374 \pm 0.0099$ &   1.017 &      80 \nl
     & WF3 & F555W & $-0.0124 \pm 0.0101$ &   1.003 &      87 \nl
     &     & F814W & $+0.0150 \pm 0.0063$ &   1.018 &     217 \nl
     & WF4 & F555W & $+0.0333 \pm 0.0119$ &   1.011 &      90 \nl
     &     & F814W & $+0.0297 \pm 0.0053$ &   1.005 &     252 \nl
G312 &  PC & F555W &        \nodata       & \nodata & \nodata \nl
     &     & F814W &        \nodata       & \nodata & \nodata \nl
     & WF2 & F555W & $-0.0344 \pm 0.0144$ &   1.008 &     186 \nl
     &     & F814W & $+0.0405 \pm 0.0087$ &   1.034 &     119 \nl
     & WF3 & F555W & $-0.0198 \pm 0.0081$ &   1.018 &     101 \nl
     &     & F814W & $-0.0030 \pm 0.0013$ &   1.015 &     125 \nl
     & WF4 & F555W & $+0.0513 \pm 0.0165$ &   1.001 &      85 \nl
     &     & F814W & $+0.0333 \pm 0.0102$ &   1.027 &      89 \nl
\enddata
\end{deluxetable}

\begin{deluxetable}{lrrrrrrrrr}
\tablewidth{0 pt}
\tablecaption{Stellar Photometry for G302 and the Surrounding Fields.\label{TABLE:G302_sample_phot}}
\tablehead
{\colhead{CCD} &
 \colhead{ID} &
 \colhead{$X$} &
 \colhead{$Y$} &
 \colhead{$V$} &
 \colhead{$\sigma_V$} &
 \colhead{$\chi_V$} &
 \colhead{$I$} &
 \colhead{$\sigma_I$} &
 \colhead{$\chi_I$}
}
\startdata
WF3 &    892 &  64.328 & 212.348 & 17.032 & 0.082 & 1.000 & 16.294 & 0.071 & 1.000 \nl
WF2 &   3440 & 466.457 & 631.376 & 17.628 & 0.084 & 0.744 & 16.365 & 0.077 & 0.727 \nl
WF3 &   2583 & 437.139 & 413.836 & 18.195 & 0.065 & 1.341 & 17.356 & 0.059 & 1.551 \nl
WF3 &   2606 & 439.952 & 415.701 & 18.454 & 0.063 & 1.462 & 17.642 & 0.064 & 1.830 \nl
WF3 &   2489 & 438.016 & 410.038 & 18.854 & 0.095 & 1.700 & 17.995 & 0.057 & 1.240 \nl
WF4 &   4532 & 300.305 & 699.237 & 19.152 & 0.031 & 0.739 & 17.418 & 0.030 & 0.877 \nl
WF3 &   2474 & 441.016 & 410.137 & 19.183 & 0.111 & 1.823 & 17.977 & 0.073 & 1.819 \nl
WF2 &    150 & 321.521 &  80.650 & 19.311 & 0.036 & 0.657 & 16.489 & 0.089 & 0.941 \nl
WF2 &   2719 & 310.787 & 523.757 & 19.669 & 0.070 & 1.499 & 18.432 & 0.063 & 1.959 \nl
WF3 &   2554 & 442.354 & 412.389 & 19.741 & 0.101 & 1.325 & 18.639 & 0.076 & 1.507 \nl
\enddata
\end{deluxetable}

\begin{deluxetable}{lrrrrrrrrr}
\tablewidth{0 pt}
\tablecaption{Stellar Photometry for G312 and the Surrounding Fields.\label{TABLE:G312_sample_phot}}
\tablehead
{\colhead{CCD} &
 \colhead{ID} &
 \colhead{$X$} &
 \colhead{$Y$} &
 \colhead{$V$} &
 \colhead{$\sigma_V$} &
 \colhead{$\chi_V$} &
 \colhead{$I$} &
 \colhead{$\sigma_I$} &
 \colhead{$\chi_I$}
}
\startdata
WF3 &    503 & 689.043 & 346.030 & 17.369 & 0.050 & 0.827 & 16.723 & 0.025 & 0.758 \nl
WF2 &   1198 & 252.106 & 634.779 & 17.758 & 0.057 & 0.745 & 16.875 & 0.055 & 0.830 \nl
WF4 &    537 & 650.877 & 360.807 & 17.934 & 0.051 & 0.982 & 17.217 & 0.026 & 0.801 \nl
WF3 &    802 & 417.826 & 417.893 & 17.983 & 0.060 & 1.509 & 17.020 & 0.061 & 1.704 \nl
WF4 &    869 & 310.743 & 507.150 & 18.569 & 0.028 & 0.630 & 17.479 & 0.026 & 0.789 \nl
WF3 &    811 & 419.875 & 419.020 & 18.662 & 0.078 & 1.841 & 17.422 & 0.053 & 1.494 \nl
WF3 &   1654 & 229.986 & 766.329 & 18.743 & 0.035 & 0.907 & 18.116 & 0.019 & 0.601 \nl
WF3 &    412 & 167.937 & 301.246 & 18.896 & 0.034 & 0.877 & 16.521 & 0.030 & 0.756 \nl
WF3 &    770 & 415.495 & 417.557 & 19.926 & 0.055 & 1.014 & 18.819 & 0.065 & 1.407 \nl
WF3 &   1451 & 676.379 & 661.313 & 20.225 & 0.038 & 0.948 & 18.614 & 0.025 & 0.736 \nl
\enddata
\end{deluxetable}

\begin{deluxetable}{lrrrrrrrrrr}
\tablewidth{0 pt}
\tablecaption{Stellar Photometry for G302.\label{TABLE:G302_GC_sample_phot}}
\tablehead
{\colhead{CCD} &
 \colhead{ID} &
 \colhead{$X$} &
 \colhead{$Y$} &
 \colhead{$r$} &
 \colhead{$V$} &
 \colhead{$\sigma_V$} &
 \colhead{$\chi_V$} &
 \colhead{$I$} &
 \colhead{$\sigma_I$} &
 \colhead{$\chi_I$}
}
\startdata
WF3 &   2583 & 437.139 & 413.836 & $0\farcs230$ & 18.195 & 0.065 & 1.341 & 17.356 & 0.059 & 1.551 \nl
WF3 &   2606 & 439.952 & 415.701 & $0\farcs332$ & 18.454 & 0.063 & 1.462 & 17.642 & 0.064 & 1.830 \nl
WF3 &   2489 & 438.016 & 410.038 & $0\farcs265$ & 18.854 & 0.095 & 1.700 & 17.995 & 0.057 & 1.240 \nl
WF3 &   2474 & 441.016 & 410.137 & $0\farcs308$ & 19.183 & 0.111 & 1.823 & 17.977 & 0.073 & 1.819 \nl
WF3 &   2554 & 442.354 & 412.389 & $0\farcs332$ & 19.741 & 0.101 & 1.325 & 18.639 & 0.076 & 1.507 \nl
WF3 &   2532 & 429.789 & 412.006 & $0\farcs921$ & 20.376 & 0.079 & 1.847 & 19.356 & 0.067 & 1.961 \nl
WF3 &   2420 & 440.762 & 402.723 & $0\farcs989$ & 20.812 & 0.071 & 1.629 & 19.648 & 0.063 & 1.828 \nl
WF3 &   2553 & 428.372 & 414.756 & $1\farcs084$ & 20.947 & 0.058 & 1.355 & 19.735 & 0.043 & 1.256 \nl
WF3 &   2460 & 418.943 & 405.789 & $2\farcs108$ & 21.148 & 0.027 & 0.679 & 19.992 & 0.029 & 0.922 \nl
WF3 &   2456 & 427.305 & 405.329 & $1\farcs368$ & 21.241 & 0.048 & 1.071 & 19.853 & 0.046 & 1.304 \nl
\enddata
\end{deluxetable}

\begin{deluxetable}{lrrrrrrrrrr}
\tablewidth{0 pt}
\tablecaption{Stellar Photometry for G312.\label{TABLE:G312_GC_sample_phot}}
\tablehead
{\colhead{CCD} &
 \colhead{ID} &
 \colhead{$X$} &
 \colhead{$Y$} &
 \colhead{$r$} &
 \colhead{$V$} &
 \colhead{$\sigma_V$} &
 \colhead{$\chi_V$} &
 \colhead{$I$} &
 \colhead{$\sigma_I$} &
 \colhead{$\chi_I$}
}
\startdata
WF3 &    802 & 417.826 & 417.893 & $0\farcs039$ & 17.983 & 0.060 & 1.509 & 17.020 & 0.061 & 1.704 \nl
WF3 &    811 & 419.875 & 419.020 & $0\farcs198$ & 18.662 & 0.078 & 1.841 & 17.422 & 0.053 & 1.494 \nl
WF3 &    770 & 415.495 & 417.557 & $0\farcs274$ & 19.926 & 0.055 & 1.014 & 18.819 & 0.065 & 1.407 \nl
WF3 &    866 & 417.573 & 427.651 & $0\farcs972$ & 22.019 & 0.086 & 1.941 & 20.250 & 0.045 & 1.217 \nl
WF3 &    783 & 429.730 & 414.665 & $1\farcs191$ & 22.086 & 0.056 & 1.330 & 20.499 & 0.050 & 1.467 \nl
WF3 &    803 & 405.224 & 418.070 & $1\farcs294$ & 22.180 & 0.053 & 1.224 & 20.275 & 0.048 & 1.413 \nl
WF3 &    841 & 396.894 & 424.414 & $2\farcs221$ & 22.310 & 0.073 & 1.267 & 20.965 & 0.059 & 1.244 \nl
WF3 &    871 & 412.104 & 427.591 & $1\farcs141$ & 22.355 & 0.072 & 1.544 & 21.183 & 0.066 & 1.812 \nl
WF3 &    915 & 418.244 & 434.008 & $1\farcs603$ & 22.645 & 0.079 & 1.777 & 21.687 & 0.068 & 1.789 \nl
WF3 &    796 & 434.824 & 416.729 & $1\farcs658$ & 22.676 & 0.041 & 0.920 & 21.468 & 0.035 & 0.953 \nl
\enddata
\end{deluxetable}

%
%

\begin{deluxetable}{ccccccc}
\tablewidth{0 pt}
\tablecaption{Photometric Uncertainties.\label{TABLE:mag_errs}}
\tablehead
{\colhead{} &
 \colhead{G302} &
 \colhead{G312} &
 \colhead{} &
 \colhead{G302} &
 \colhead{G312} \nl
 \colhead{$V$} &
 \colhead{$\sigma_V$} &
 \colhead{$\sigma_V$} &
 \colhead{$I$} &
 \colhead{$\sigma_I$} &
 \colhead{$\sigma_I$}
}
\startdata
 20.25 & \nodata & \nodata &      20.25 &   0.031 & \nodata \nl
 20.75 & \nodata & \nodata &      20.75 &   0.026 &   0.023 \nl
 21.25 & \nodata & \nodata &      21.25 &   0.031 &   0.028 \nl
 21.75 &   0.038 & \nodata &      21.75 &   0.036 &   0.033 \nl
 22.25 &   0.037 & \nodata &      22.25 &   0.038 &   0.039 \nl
 22.75 &   0.043 &   0.046 &      22.75 &   0.046 &   0.043 \nl
 23.25 &   0.049 &   0.049 &      23.25 &   0.051 &   0.049 \nl
 23.75 &   0.055 &   0.051 &      23.75 &   0.067 &   0.063 \nl
 24.25 &   0.065 &   0.062 &      24.25 &   0.081 &   0.077 \nl
 24.75 &   0.079 &   0.077 &      24.75 &   0.118 &   0.099 \nl
 25.25 &   0.094 &   0.090 &      25.25 &   0.145 &   0.153 \nl
 25.75 &   0.129 &   0.121 &      25.75 &   0.217 &   0.222 \nl
 26.25 &   0.176 &   0.182 &      26.25 &   0.333 &   0.314 \nl
 26.75 &   0.262 &   0.271 &      26.75 &   0.570 &   0.457 \nl
 27.25 &   0.356 &   0.310 &      27.25 & \nodata & \nodata \nl
 27.75 & \nodata & \nodata &      27.75 & \nodata & \nodata \nl
 28.25 & \nodata & \nodata &      28.25 & \nodata & \nodata \nl
\enddata
\end{deluxetable}

%
%

\begin{deluxetable}{rrrrrrrrrr}
\tablewidth{0 pt}
\tablecaption{Magnitude shifts in RGB artificial star data.\label{TABLE:art_shift_RGB}}
\tablehead
{\colhead{$V$} &
 \colhead{$[V]$} &
 \colhead{$[\sigma_V]$} &
 \colhead{$[\delta_V]$} &
 \colhead{$\hat\sigma_V$} &
 \colhead{$[(V-I)_0]$} &
 \colhead{$[\sigma_{(V-I)_0}]$} &
 \colhead{$[\delta_{(V-I)_0}]$} &
 \colhead{$\hat\sigma_{(V-I)_0}$} &
 \colhead{$N$}
}
\startdata
 22.5--23.0  &  22.947 &  0.058 & $-0.008$ &  0.039 &  1.536 &  0.074 &  $0.002$ &  0.049 &    2 \nl
 23.0--23.5  &  23.361 &  0.065 & $-0.030$ &  0.061 &  1.382 &  0.082 & $-0.013$ &  0.028 &   19 \nl
 23.5--24.0  &  23.800 &  0.064 & $-0.006$ &  0.056 &  1.237 &  0.089 & $-0.010$ &  0.062 &   31 \nl
 24.0--24.5  &  24.262 &  0.081 &  $0.024$ &  0.055 &  1.081 &  0.102 &  $0.009$ &  0.064 &   55 \nl
 24.5--25.0  &  24.781 &  0.087 &  $0.023$ &  0.085 &  1.007 &  0.115 &  $0.028$ &  0.087 &   78 \nl
 25.0--25.5  &  25.251 &  0.104 &  $0.041$ &  0.098 &  0.918 &  0.138 & $-0.006$ &  0.094 &  110 \nl
 25.5--26.0  &  25.752 &  0.128 &  $0.043$ &  0.176 &  0.890 &  0.177 & $-0.009$ &  0.187 &  116 \nl
 26.0--26.5  &  26.235 &  0.179 &  $0.068$ &  0.277 &  0.881 &  0.244 &  $0.004$ &  0.192 &   88 \nl
 26.5--27.0  &  26.654 &  0.232 &  $0.138$ &  0.412 &  0.873 &  0.319 &  $0.042$ &  0.184 &   33 \nl
 27.0--27.5  &  27.076 &  0.308 &  $0.347$ &  0.052 &  0.826 &  0.415 & $-0.011$ &  0.169 &    3 \nl
\enddata
\end{deluxetable}

%
%

\begin{deluxetable}{rrrrrrrrrr}
\tablewidth{0 pt}
\tablecaption{Magnitude shifts in HB artificial star data.\label{TABLE:art_shift_HB}}
\tablehead
{\colhead{$V$} &
 \colhead{$[V]$} &
 \colhead{$[\sigma_V]$} &
 \colhead{$[\delta_V]$} &
 \colhead{$\hat\sigma_V$} &
 \colhead{$[(V-I)_0]$} &
 \colhead{$[\sigma_{(V-I)_0}]$} &
 \colhead{$[\delta_{(V-I)_0}]$} &
 \colhead{$\hat\sigma_{(V-I)_0}$} &
 \colhead{$N$}
}
\startdata
 24.0--24.5  &  24.384 &  0.073 & $-0.744$ &  0.100 &  0.623 &  0.106 &  $0.099$ &  0.159 &   12 \nl
 24.5--25.0  &  24.911 &  0.086 & $-0.194$ &  0.123 &  0.429 &  0.134 &  $0.010$ &  0.181 &   94 \nl
 25.0--25.5  &  25.182 &  0.102 &  $0.055$ &  0.090 &  0.537 &  0.153 & $-0.020$ &  0.136 &  823 \nl
 25.5--26.0  &  25.615 &  0.214 &  $0.486$ &  0.200 &  0.945 &  0.236 &  $0.400$ &  0.293 &   12 \nl
 26.0--26.5  &  26.054 &  0.246 &  $0.890$ &  0.000 &  1.569 &  0.274 &  $0.856$ &  0.000 &    1 \nl
\enddata
\end{deluxetable}

%
%

\begin{deluxetable}{lccccrrc}
\tablewidth{0 pt}
\tablecaption{The GC Data.\label{TABLE:vi-feh}}
\tablehead
{\colhead{Cluster} &
 \colhead{$(V-I)_{0,-3}$} &
 \colhead{$\FeH_{\rm S}$} &
 \colhead{$\FeH_{\rm CMD}$} &
 \colhead{$E_{V-I}$} &
 \colhead{$R_{\rm M31}$} &
 \colhead{$Y$} &
 \colhead{Reference}
}
\startdata
G1   & $1.47 \pm 0.06$ & $-1.08 \pm 0.09$ & $-0.65 \pm 0.10$ & $0.10 \pm 0.03$ & 152\farcm3 & $+29\farcm1$ & 4 \nl
G11  & $1.22 \pm 0.04$ & $-1.89 \pm 0.17$ & $-1.7  \pm 0.20$ & $0.10 \pm 0.03$ &  75\farcm7 & $+43\farcm6$ & 2 \nl
G58  & $1.80 \pm 0.05$ & $-0.57 \pm 0.15$ & $-0.57 \pm 0.15$ & $0.14 \pm 0.04$ &  28\farcm2 & $+27\farcm3$ & 1 \nl
G105 & $1.31 \pm 0.02$ & $-1.49 \pm 0.17$ & $-1.49 \pm 0.17$ & $0.08 \pm 0.02$ &  64\farcm8 & $-29\farcm9$ & 1 \nl
G108 & $1.62 \pm 0.05$ & $-0.94 \pm 0.27$ & $-0.80 \pm 0.10$ & $0.15 \pm 0.04$ &  20\farcm8 & $+19\farcm7$ & 1 \nl
G219 & $1.20 \pm 0.02$ & $-1.83 \pm 0.22$ & $-2.04 \pm 0.22$ & $0.08 \pm 0.02$ &  87\farcm2 & $-58\farcm8$ & 1 \nl
G302 & $1.24 \pm 0.04$ & $-1.76 \pm 0.18$ & $-1.85 \pm 0.12$ & $0.10 \pm 0.03$ &  32\farcm1 & $-30\farcm4$ & 3 \nl
G312 & $1.88 \pm 0.07$ & $-0.70 \pm 0.35$ & $-0.56 \pm 0.03$ & $0.10 \pm 0.03$ &  49\farcm8 & $-49\farcm7$ & 3 \nl
G319 & $1.70 \pm 0.17$ & $-0.66 \pm 0.22$ & $-0.6  \pm 0.90$ & $0.10 \pm 0.03$ &  72\farcm1 & $-69\farcm0$ & 2 \nl
G323 & $1.17 \pm 0.04$ & $-1.96 \pm 0.29$ & $-2.0  \pm 0.20$ & $0.10 \pm 0.03$ &  53\farcm8 & $-53\farcm8$ & 2 \nl
G327 & $1.31 \pm 0.06$ & $-1.78 \pm 0.11$ & $-1.3  \pm 0.30$ & $0.10 \pm 0.03$ &  99\farcm7 & $+19\farcm9$ & 2 \nl
G352 & $1.84 \pm 0.06$ & $-0.85 \pm 0.33$ & $-0.5  \pm 0.10$ & $0.10 \pm 0.03$ &  87\farcm1 & $-49\farcm6$ & 2 \nl
\enddata

\tablerefs{(1) Ajhar \etal \markcite{AG96}1996;
(2) Couture \etal \markcite{CR95}1995;
(3) this work;
(4) Rich \etal \markcite{RM96}1996.
}

\end{deluxetable}

%
%

\begin{deluxetable}{rrrrr}
\tablewidth{0 pt}
\tablecaption{$V$-Band Luminosity Functions for G302 and G312.\label{TABLE:lfs}}
\tablehead
{\colhead{$V$} &
 \colhead{$n_{\rm G302}(V)$} &
 \colhead{$\phi_{\rm G302}(V)$} &
 \colhead{$n_{\rm G312}(V)$} &
 \colhead{$\phi_{\rm G312}(V)$}
}
\startdata
 21.75 &    3 &    2.719 &    0 &    0.000 \nl
 22.25 &   17 &   16.184 &    0 &    0.000 \nl
 22.75 &   26 &   22.139 &    2 &    0.725 \nl
 23.25 &   32 &   20.397 &   19 &   16.635 \nl
 23.75 &   46 &   34.960 &   12 &    4.148 \nl
 24.25 &   58 &   61.228 &   25 &   26.724 \nl
 24.75 &  125 &  166.622 &   48 &   57.009 \nl
 25.25 &  116 &   43.772 &   85 &  139.459 \nl
 25.75 &   62 &    0.000 &   36 &   61.921 \nl
 26.25 &   50 &  142.707 &   26 &  157.769 \nl
 26.75 &   25 &    0.000 &   13 &  313.156 \nl
 27.25 &    4 &    0.000 &    3 &  216.386 \nl
\enddata
\end{deluxetable}

%
%

\begin{deluxetable}{ccrrrrrrrr}
\tablewidth{0 pt}
\tablecaption{Best-Fitting Michie--King Models.\label{TABLE:1d_models}}
\tablehead
{\colhead{Cluster} &
 \colhead{Filter} &
 \colhead{$W_0$} &
 \colhead{$r_c$} &
 \colhead{$r_t$} &
 \colhead{$c$} &
 \colhead{$r_a$} &
 \colhead{$r_h$} &
 \colhead{$\chi_\nu^2$} &
 \colhead{$\nu$}
}
\startdata
G302 & $V$ & 7.56 & $0\farcs20$ &  $9\farcs92$ & 1.70 &   $+\infty$   & $0\farcs52$ & 1.827 & 49 \nl
     &     & 7.61 & $0\farcs20$ & $10\farcs12$ & 1.71 & $282\farcs40$ & $0\farcs54$ & 1.855 & 48 \nl
     & $I$ & 7.65 & $0\farcs20$ & $10\farcs49$ & 1.73 &   $+\infty$   & $0\farcs56$ & 0.846 & 51 \nl
     &     & 7.47 & $0\farcs21$ &  $9\farcs80$ & 1.67 &  $20\farcs06$ & $0\farcs49$ & 0.906 & 48 \nl
G312 & $V$ & 7.57 & $0\farcs19$ &  $9\farcs51$ & 1.70 &   $+\infty$   & $0\farcs53$ & 0.839 & 53 \nl
     &     & 7.48 & $0\farcs20$ &  $9\farcs16$ & 1.67 & $492\farcs13$ & $0\farcs49$ & 0.882 & 52 \nl
     & $I$ & 7.46 & $0\farcs21$ &  $9\farcs64$ & 1.67 &   $+\infty$   & $0\farcs49$ & 0.458 & 53 \nl
     &     & 7.47 & $0\farcs21$ &  $9\farcs70$ & 1.67 & $381\farcs88$ & $0\farcs49$ & 0.467 & 52 \nl
\enddata
\end{deluxetable}

%
%

\begin{deluxetable}{rccrcl}
\tablewidth{0 pt}
\tablecaption{Ellipticities and position angles.\label{TABLE:shapes}}
\tablehead
{\colhead{Cluster} &
 \colhead{Filter} &
 \colhead{$\overline{\epsilon}$} &
 \colhead{$\overline{\theta_0}$} &
 \colhead{$N$}
}
\startdata
G302 & $V$ & $0.194 \pm 0.011$ & $-71\arcdeg \pm 2\arcdeg$ & 24 \nl
     & $I$ & $0.195 \pm 0.012$ & $-87\arcdeg \pm 2\arcdeg$ & 24 \nl
G312 & $V$ & $0.072 \pm 0.015$ &  $77\arcdeg \pm 5\arcdeg$ & 24 \nl
     & $I$ & $0.058 \pm 0.014$ &  $77\arcdeg \pm 6\arcdeg$ & 24 \nl
\enddata
\end{deluxetable}


\newpage

\clearpage

\clearpage

\end{document}